\documentclass[aps,prd,floatfix,preprintnumbers,superscriptaddress,notitlepage,fleqn]{revtex4-1}
\usepackage{dcolumn}
\usepackage{bm}
\usepackage{color}
\usepackage{amssymb}
\usepackage{amsmath}
\usepackage{graphicx}
\usepackage{amsfonts}
\usepackage{slashed}
\usepackage{float}
\usepackage[colorlinks = true,
            linkcolor = blue,
            urlcolor  = blue,
            citecolor = blue,
            anchorcolor = blue]{hyperref}
\usepackage{array}
\allowdisplaybreaks
\usepackage{dcolumn}
\usepackage{epsf}
\usepackage{slashed}

\usepackage{color}
\definecolor{gray}{rgb}{0.6,0.6,0.6}
\definecolor{darkgreen}{rgb}{0.0, 0.545098, 0.0}
\definecolor{darkblue}{rgb}{0.0, 0.0, 0.545098}

\mathchardef\mhyphen="2D 

\begin{document}
\title{Tau Polarization and Correlated Decays in Neutrino Experiments}
\author{{Joshua} Isaacson}
\affiliation{Theory Division, Fermi National Accelerator Laboratory, P.O. Box 500, Batavia, IL 60510, USA}
\author{{Stefan} H\"oche}
\affiliation{Theory Division, Fermi National Accelerator Laboratory, P.O. Box 500, Batavia, IL 60510, USA}
\author{Frank Siegert}
\affiliation{Institut f{\"u}r Kern- und Teilchenphysik, TU Dresden, 01069 Dresden, Germany}
\author{{Sherry} Wang}
\affiliation{Northwestern University, Department of Physics \& Astronomy, 2145 Sheridan Road, Evanston, IL 60208, USA}
\preprint{FERMILAB-PUB-23-106-T, MCNET-23-04}

\begin{abstract}
We present the first fully differential predictions for tau neutrino scattering in the energy region
relevant to the DUNE experiment, including all spin correlations and all tau lepton decay channels.
The calculation is performed using a generic interface between the neutrino event generator Achilles
and the publicly available, general-purpose collider event simulation framework Sherpa.
\end{abstract}
\maketitle
\section{Introduction}\label{sec:introduction}

The tau neutrino is commonly considered to be the least well known elementary particle.
The first experimental direct evidence for tau neutrinos was provided about two decades ago
by the DONuT experiment~\cite{DONUT:2000fbd}. Major limitations on the dataset came from 
a small cross section, the large mass of the tau lepton, and the large irreducible backgrounds.
As of today, there are still very few positively identified tau neutrino events
from collider based sources, with 9 detected by DONuT~\cite{DONuT:2007bsg}, 
and 10 detected by OPERA~\cite{OPERA:2018nar}. The SuperK~\cite{Super-Kamiokande:2017edb}, and
IceCube~\cite{IceCube:2019dqi,IceCube:2020fpi} experiments have identified 291 and 1806
tau neutrino candidates
from atmospheric and astrophysical sources. New experiments are expected to come online soon, 
among them DUNE~\cite{DeGouvea:2019kea,DUNE:2020ypp} and the IceCube upgrade~\cite{Ishihara:2019aao},
which will improve the precision on the $\nu_\mu \to \nu_\tau$ appearance measurement.
The forward physics facility~\cite{Feng:2022inv} will use the large forward charm 
production rate at the LHC to perform precision studies with collider neutrinos.
Ultra-high energy neutrino telescopes will set limits on $\nu_\tau$ self-interactions (which
are currently unconstrained~\cite{Esteban:2021tub}) and flavor ratios (which are an important observable to constrain new physics~\cite{Arguelles:2019rbn}).
With all of these novel experiments, the
tau neutrino dataset is expected to grow quickly in the coming years, creating new opportunities
for measurements and searches for physics beyond the standard neutrino paradigm~\cite{MammenAbraham:2022xoc}.

DUNE is especially important to the tau neutrino program, since it will be the only accelerator
based experiment able to collect and accurately reconstruct a sample of oscillated $\nu_\tau$
charged current (CC) events, with about 130 $\nu_\tau$ CC events per year in CP-optimized neutrino mode,
30 $\bar{\nu}_\tau$ events per year in CP-optimized antineutrino mode and about 800 $\nu_\tau$ 
CC events per year in tau-optimized neutrino mode~\cite{MammenAbraham:2022xoc}.
To make the most of these events, accurate theory predictions are required. One key observation
to help separate the signal from the irreducible background is the fact that the tau is polarized, leading
to correlations in the outgoing pions.
However, the produced outgoing tau lepton is not fully polarized for DUNE
energies~\cite{Sobczyk:2019urm,Hernandez:2022nmp}. Additionally, the cross section is dominated
by quasielastic and resonance scattering. Computational tools that model both the intricate aspects
of nuclear physics involved in $\nu$-nucleus interactions and the effects of polarized scattering 
and decay are vital for experimental success~\cite{Machado:2020yxl}. However, the existing neutrino 
event generators GENIE~\cite{Andreopoulos:2009rq}, NuWro~\cite{Golan:2012rfa}, NEUT~\cite{Hayato:2021heg},
and GiBUU~\cite{Leitner:2006ww} generate $\nu_\tau$ interactions in the same manner 
as $\nu_e$ and $\nu_\mu$ events. They then assume that the outgoing $\tau$ is purely left-handed
and simulate its decay with the help of TAUOLA~\cite{Chrzaszcz:2016fte}.
We will address this shortcoming by constructing an event generator based on a state-of-the art
nuclear physics model, in combination with a general-purpose tau decay simulation including
spin correlations between the production and all subsequent decays.

Various theoretical calculations have also addressed nuclear effects on the polarization of the tau
in neutrino scattering. However, the previous works either do not include tau decays~\cite{Sobczyk:2019urm},
or they only include the one-body decay of the tau (\textit{i.e.}\ $\tau^- \to \nu_\tau \pi^-$)~\cite{Hernandez:2022nmp}.
They demonstrate the dependence of the nuclear effects on the
polarization and the impact on observables, respectively. Here, we extend these
studies to include all possible decay channels of the tau, while maintaining complete polarization
information, and we provide a publicly available simulation package to generate fully differential
final states. The calculation is performed using Achilles~\cite{Isaacson:2022cwh} to handle 
the nuclear physics effects and Sherpa~\cite{Gleisberg:2003xi,Gleisberg:2008ta,Sherpa:2019gpd}
to perform the leptonic calculation and the decay of the tau. This interface extends
the one developed in Ref.~\cite{Isaacson:2021xty}, which also allows to perform the calculation
in nearly arbitrary new physics models by means of FeynRules~\cite{Christensen:2008py,Alloul:2013bka}.

The outline of this paper is as follows. In Sec.~\ref{sec:tau_overview}, we review analytic results on the 
production and decay of the tau, with a focus on the effects of nuclear physics and the high energy limit.
The implementation of tau decays within the Sherpa framework and the interface between Achilles and
Sherpa is described in Sec.~\ref{sec:Sherpa}. Comparisons for purely left-handed and the correct 
polarization is shown for various monochromatic neutrino beam energies as well as for a realistic
tau-optimized DUNE neutrino flux in Sec.~\ref{sec:Results}.

\section{Polarization in tau lepton production and decay}
\label{sec:tau_overview}
This section provides a brief overview of the main analytic results on the effect of polarization 
in $\tau$ decays and production. The collinear limit, which provides both theoretical insight and a useful 
benchmark for the validation of Monte-Carlo simulations, is discussed in some detail. Furthermore, the dependency
 of the polarization of the $\tau$ on the hadronic tensor is reviewed. 

\subsection{Tau Decays in the Collinear Limit}
\label{sec:collinear_limit}
The dominant decay channels of the $\tau$ are into a single pion, leptons, or into a vector meson resonance.
In these channels, ignoring the decays of the vector mesons, the distribution of the final state momenta
can be determined in the collinear limit (\textit{i.e.} $p_\tau \to \infty$). These results are useful
for the validation of more detailed theoretical predictions.

\begin{figure}
    \centering
    \includegraphics[width=0.5\textwidth]{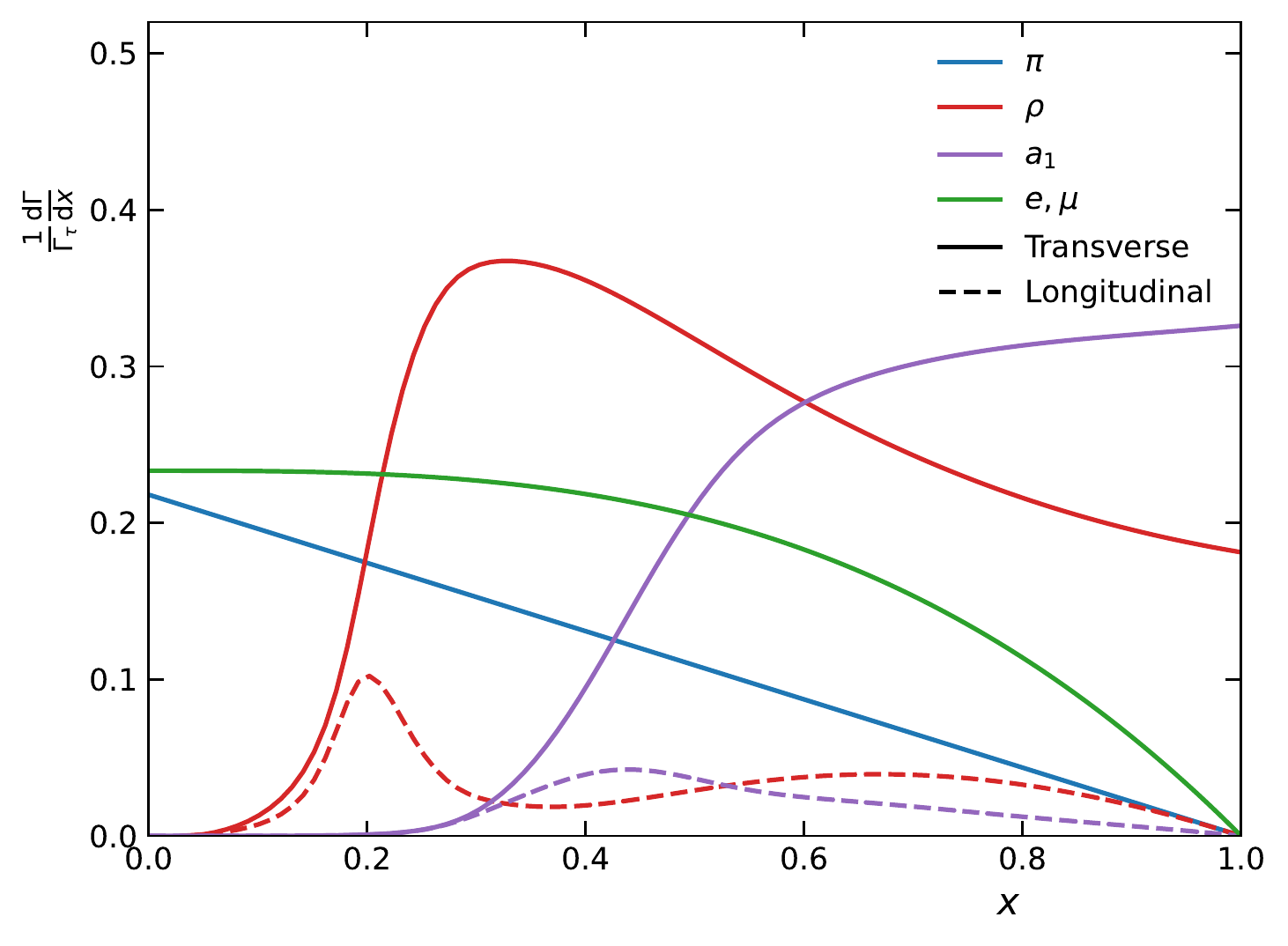}
    \caption{The decay distributions vs the fractional momentum of a given particle to the $\tau$ momentum
    for a left-handed $\tau^-$ in the collinear limit going
    to single pions (blue), $\rho$ mesons (red), $a_1$ mesons (purple), or leptons (green). The vector
    mesons ($\rho$ and $a_1$) can be either transversely polarized (solid lines) or
    longitudinally polarized (dashed lines).
    Additionally, the vector mesons are not stable and the effect of their widths are included, which
    is set to 0.1474 GeV and 0.420 GeV for the $\rho$ and $a_1$ respectively.}
    \label{fig:xpi_collinear}
\end{figure}

The rate of the $\tau^{\mp}\rightarrow \pi^{\mp} \nu_\tau$ decay in the rest frame of the tau
is given as~\cite{Bullock:1992yt}
\begin{equation}\label{eq:cospi}
    \frac{1}{\Gamma_\tau} \frac{{\rm d} \Gamma_\pi}{{\rm d}\cos\theta_\pi} =
    \frac{1}{2} B_\pi\left(1 \pm P_\tau \cos\theta_\pi\right)\,,
\end{equation}
where $B_\pi$ is the branching fraction of $\tau\rightarrow\pi\nu_\tau$, $P_\tau$ is the polarization of 
the $\tau$, and $\theta_\pi$ is the angle between the pion momentum and the tau spin axis, which 
coincides with the $\tau$ momentum in the lab frame. For a purely right-(left)-handed $\tau^-$, the 
polarization is $P_\tau = +1 (-1)$. In terms of the momentum fraction, $x_\pi = E_\pi/E_\tau$, the 
polar angle is given as
\begin{equation}\label{eq:cosz}
    \cos\theta_\pi = \frac{2 x_\pi - 1 - a^2}{\beta (1-a^2)}\,,
\end{equation}
where $a=m_\pi/m_\tau$ and $\beta$ is the velocity of the $\tau$. In the collinear limit, $\beta 
\rightarrow 1$, and making the approximation $a=0$, one obtains
\begin{equation}\label{eq:xpi}
    \frac{1}{\Gamma_\tau} \frac{{\rm d} \Gamma_\pi}{{\rm d}x_\pi} =
    B_\pi \left(1\pm P_\tau\left(2 x_\pi -1\right)\right)\,.
\end{equation}
In this limit, we obtain the prediction for the differential decay rate 
shown in Fig.~\ref{fig:xpi_collinear}.

Additionally, for the case of leptonic decays in the collinear and massless limit ($m_e = m_\mu = 0$) the
tau decay to leptons is the same for electrons and muons. The differential decay rate is given by~\cite{Bullock:1992yt}
\begin{equation}
    \frac{1}{\Gamma_\tau}\frac{{\rm d} \Gamma_\ell}{{\rm d}x_\ell} =
    \frac{1}{3}B_\ell (1-x_\ell)\left(\left(5+5x_\ell-4x_\ell^2\right)\mp\left(1+x_\ell-8x_\ell^2\right)\right)\,,
\end{equation}
where $x_\ell = p_\ell / p_\tau$, and $B_\ell$ is the branching ratio into a given lepton.
The rate for leptons is shown in Fig.~\ref{fig:xpi_collinear}.

Similarly, the decays for the vector meson decay modes $\tau \to v \nu_\tau$, with $v = \rho$ or $a_1$
are calculated in Ref.~\cite{Bullock:1992yt} and the results are reproduced here for convenience.
The mesons are separated into the transverse and longitudinal components in the calculation, since the decays
$\rho \to 2\pi$ and $a_1 \to 3\pi$ depend on the polarization of the vector mesons. 
The angular distribution in the rest frame of the tau is given as:
\begin{align}
    \frac{1}{\Gamma_\tau} \frac{{\rm d} \Gamma^T_v}{{\rm d}\cos\theta_v} &=
    B_v\frac{m_v^2}{m_\tau^2+2m_v^2}\left(1\mp P_\tau \cos\theta_v\right)\,, \label{eq:vector_meson_T} \\
    \frac{1}{\Gamma_\tau} \frac{{\rm d} \Gamma^L_v}{{\rm d}\cos\theta_v} &=
    B_v\frac{\frac{1}{2}m_v^2}{m_\tau^2+2m_v^2} \left(1\pm P_\tau \cos\theta_v\right)\,,
    \label{eq:vector_meson_L}
\end{align}
where again $v = \rho$ or $a_1$, $B_v$ is the branching ratio for $\tau^\mp \to v^\mp \nu_\tau$,
and $\theta_v$ is the same angle defined in the pion case.
It is important to note that for the case of the longitudinal state the polarization dependence
is the same as Eq.~\eqref{eq:cospi}, while for the transverse state the polarization enters with
the opposite sign. Therefore, if the polarization of the vector meson is not measured, then
Eqs.~\eqref{eq:vector_meson_T} and~\eqref{eq:vector_meson_L} need to be averaged. This suppresses the
sensitivity to the polarization of the tau by a factor of $(m_\tau^2 - 2m_v^2)/(m_\tau^2+2m_v^2)$,
which is about 0.46 for the case of the $\rho$ and approximately 0.02 for the case of the $a_1$ meson.

In the case of the vector mesons, care has to be taken when boosting to the lab frame since the 
polarizations are not summed over. First, a Wigner rotation~\cite{Martin:1970hmp} is used to
align the spin axis. The angle of rotation is given in the collinear limit by~\cite{Bullock:1992yt}
\begin{equation}
    \cos\omega = \frac{1-a^2+(1+a^2)\cos\theta}{1+a^2+(1-a^2)\cos\theta}\,,
\end{equation}
where $a=m_v/m_\tau$. Rewriting in terms of the
momentum fraction ($x_v=E_v/E_\tau$), the decay distributions can be expressed as
\begin{align}
    \frac{1}{\Gamma_\tau}\frac{{\rm d}\Gamma_v}{{\rm d}x_v} &= B_v H_v^\alpha(x_v, m_v^2)\,,
\end{align}
where $\alpha = T$ or $L$ and the expressions for $H_v^{T,L}$ are given in Eqs.~(2.16) and~(2.17) of 
Ref.~\cite{Bullock:1992yt} respectively. The results for the decay distribution including the width for
a left-handed $\tau^-$ decay are shown in Fig.~\ref{fig:xpi_collinear}.
These distributions provide the main analytic benchmark points for tests of our Monte-Carlo implementation.

\subsection{Production of the tau lepton}
The unpolarized differential cross-section for CC interaction $\nu_\tau A \to \tau^- X$ can be expressed
as the product of a leptonic and hadronic tensor as shown in Ref.~\cite{Isaacson:2021xty}.
In the case of a massive lepton, there are six nuclear
structure functions that appear in the hadronic tensor with an associated Lorentz structure~\cite{Hernandez:2022nmp}
\begin{equation}
    \frac{W^{\mu\nu}}{2M_A} = -g^{\mu\nu}W_1 + \frac{P^\mu P^\nu}{M_A^2} W_2 
    + i\frac{\epsilon^{\mu\nu\gamma\delta}P_{\gamma}q_{\delta}}{2M_A^2}W_3+\frac{q^\mu q^\nu}{M_A^2}W_4
    + \frac{P^\mu q^\nu + P^\nu q^\mu}{2M_A^2}W_5 + i\frac{P^\mu q^\nu - P^\nu q^\mu}{2M_A^2}W_6\,,
\end{equation}
where $M_A$ is the mass of the nucleus, $P^\mu$ is the initial momentum of the nucleus, $q^\mu$ is the momentum
transfer, and $\epsilon^{\mu\nu\gamma\delta}$ is the fully anti-symmetric tensor with $\epsilon^{0123} = +1$.

The unpolarized, longitudinal, and transverse components for the production of the $\tau$ can be expressed as
different linear combinations of the hadronic structure functions. These are given in Eqs.~(2),~(5), and~(6) of
Ref.~\cite{Valverde:2006yi} and are reproduced here for completeness.
\begin{align}
    F &= \left(2 W_1 + \frac{m_l^2}{M_A^2} W_4\right)\left(E_l-|\vec{p}_l|\cos\theta\right)
      + W_2\left(E_l+|\vec{p}_l|\cos\theta\right)- W_5\frac{m_l^2}{M_A} \\
      &\mp\frac{W_3}{M_A}\left(E_\nu E_l + |\vec{p}_l|^2 
                               - \left(E_\nu + E_l\right)|\vec{p}_l|\cos\theta\right)\,, \nonumber \\
    P_L &= \mp\left(\left(2W_1-\frac{m_l^2}{M_A^2}W_4\right)\left(|\vec{p}_l|-E_l\cos\theta\right)
      + W_2\left(|\vec{p}_l|+E_l\cos\theta\right)- W_5\frac{m_l^2}{M_A}\cos\theta\right. \\
      &\left.\mp\frac{W_3}{M_A}\left(\left(E_\nu + E_l\right)|\vec{p}_l|
                               -\left(E_\nu E_l + |\vec{p}_l|^2\right)\cos\theta\right)\right) / F\,, \nonumber \\
    P_T &= \mp m_l \sin\theta\left(2W_1 - W_2-\frac{m_l^2}{M_A^2}W_4+W_5\frac{E_l}{M_A}\mp W_3\frac{E_\nu}{M_A}\right) / F\,,
\end{align}
where $E_l, m_l, \vec{p}_l$ is the outgoing lepton energy, mass, and three momentum, respectively. Additionally,
$\cos\theta$ is the outgoing lepton angle with respect to the neutrino direction and $E_\nu$ is the energy of the
incoming neutrino. It is important to note that the above equations are insensitive to the $W_6$ structure function.
Furthermore, the structure functions $W_4$ and $W_5$ are proportional to the mass of the lepton and are only
weakly constrained due to the limited statistics on tau-neutrino-nucleus scattering. The limits DUNE
can set on the structure functions, from using the combination of both inclusive and differential
rates, would provide valuable constraints on nuclear models used to describe neutrino-nucleus
interactions~\cite{Hernandez:2022nmp}. Additionally, DUNE will be the first experiment to provide measurements
of the $W_4$ and $W_5$ structure functions in the quasielastic region, directly testing the partially conserved
axial current and the pion-pole dominance ansatz~\cite{MammenAbraham:2022xoc}.

\section{Monte-Carlo Simulation}
\label{sec:Sherpa}
In this section we will review our approach to the simulation of the scattering and decay processes. 
We make use of the fact that the reaction factorizes into a leptonic and a hadronic component. 
We employ the neutrino event generator Achilles~\cite{Isaacson:2022cwh} to handle the nuclear physics
effects and the general-purpose event generation framework Sherpa~\cite{Gleisberg:2003xi,
  Gleisberg:2008ta,Sherpa:2019gpd} to perform the leptonic calculation and the decay of the tau.
The Sherpa framework includes two modules to simulate decays of unstable particles: one for prompt decays of particles produced in the hard scattering process perturbatively, and one for the decay of hadrons produced during the hadronization stage of event generation. The tau lepton plays a special role, as it can be produced in the hard scattering process, but is the only lepton that can decay into hadrons. For a good modeling of tau decays and also for the hadronic decay modes we thus employ the hadron decay module~\cite{Laubrich:2006aa,Siegert:2006xx}. It enables us to use elaborate form factor models, accurate branching fractions for individual hadronic final states, and spin correlation effects for the decaying tau lepton. We briefly describe these features in the following.

\subsection{The decay cascade}
With the observed tau decay channels in the PDG~\cite{ParticleDataGroup:2022pth} accounting for roughly 100\% of the tau width, we use these values directly for the simulation by choosing a decay channel according to the measured branching fractions. This can include fully leptonic decay channels as well as decays into up to 6 hadrons.

Matrix elements are used to simulate the kinematical distribution of the decay in phase space. In the case of weak tau decays, these matrix elements will always contain a leptonic current $L^{(\tau\to\nu_\tau)}_\mu$ involving the $\tau$ and $\nu_\tau$ leptons, and a second current involving either another lepton pair or hadronic decay products. Due to the low tau mass and the low related momentum transfer $Q^2 \ll m_W^2$, the $W$ propagator between these currents can be integrated out into the Fermi constant
\begin{equation}
    \label{eq:me}
    \mathcal{M} = \frac{G_F}{\sqrt{2}} L^{(\tau\to\nu_\tau)}_\mu \, J^\mu\;.
\end{equation}
For currents $J^\mu$ involving hadronic final states, these matrix elements can not be derived from first principles, but are instead based on the spin of the involved particles and include form factors to account for bound-state effects and hadronic resonances within the hadronic current in particular.

\subsection{Form factor models in hadronic currents}
While the current for the production of a single meson is trivial and determined fully by the meson's decay constant, the currents in multiple-meson production can contain resonance structures. For example, in the production of pions and kaons the main effects stem from intermediate vector mesons with a short life time, like $\rho$ or $K^*$. In the Sherpa simulation, the currents are thus supplemented with form factors that parametrize these effects using one of two approaches~\cite{Laubrich:2006aa}.

The K\"uhn-Santamaria (KS) model~\cite{Kuhn:1990ad} is a relatively simple approach modeling resonances based on their Breit-Wigner distribution. Multiple resonances can contribute to the same current and are weighted with parameters that are fit to experimental data. The width in the Breit-Wigner distribution is calculated as a function of the momentum transfer.

Another approach for the form factor is based on Resonance Chiral Theory (R$\chi$T)~\cite{Ecker:1988te}, an extension of chiral perturbation theory to higher energies where resonances become relevant. Also here an energy-dependent width is used for the implementation of the resonances. This form factor model is superior for final states dominated by one resonance but cannot model multiple resonances. It will thus yield significant differences with respect to the KS model for any channel where the lower-lying resonances are kinematically suppressed, e.g.\ two-kaon production.

\subsection{Spin correlations}\label{sec:SpinDecays}
The implementation of spin correlations in the Monte-Carlo simulation of particle decays is described in detail in Ref.~\cite{Richardson:2001df}. This algorithm uses spin-density matrices to properly track polarization information through the decays. Here we summarize only its main features.
Firstly, the matrix element is evaluated for all possible spin states for the initial and final state
($\mathcal{M}_{\kappa_1\kappa_2;\lambda_1\ldots\lambda_n}$),
where $\kappa_i$ is the spin of the spin of the $i$th incoming particle and $\lambda_j$ is the spin of the $j$th outgoing particle in a $2\rightarrow n$ scattering process. 
The matrix element squared involved in the calculation of the differential cross-section can be obtained as
\begin{equation}
    \label{eq:spin_decay}
    \rho^1_{\kappa_1\kappa_1'}\rho^2_{\kappa_2\kappa_2'}\mathcal{M}_{\kappa_1\kappa_2;\lambda_1\ldots\lambda_n}\mathcal{M}^*_{\kappa_1'\kappa_2';\lambda_1'\ldots\lambda_n'}\prod_{i=1,n}D^i_{\lambda_i\lambda_i'}\,,
\end{equation}
where $\rho^i_{\kappa_i\kappa_i'}$ is the spin density matrix for the incoming particles and $D^i_{\lambda_i\lambda_i'}$ is the spin-dependent decay matrix for the outgoing particles. Before any decays occur, the decay matrix is given as $D^i_{\lambda_i\lambda_i'}=\delta_{\lambda_i\lambda_i'}$ and the spin density matrix is given as $\rho^i_{\kappa_i\kappa_i'} = \frac{1}{2}\delta_{\kappa_i\kappa_i'}$ for unpolarized incoming particles.
Secondly, one of the unstable final state particles is selected at random to decay and the spin density matrix is calculated as
\begin{equation}
    \rho_{\lambda_j\lambda_j'} = \frac{1}{N_p} 
    \rho^1_{\kappa_1\kappa_1'}\rho^2_{\kappa_2\kappa_2'}\mathcal{M}_{\kappa_1\kappa_2;\lambda_1\ldots\lambda_n}\mathcal{M}^*_{\kappa_1'\kappa_2';\lambda_1'\ldots\lambda_n'}\prod_{i\neq j}D^i_{\lambda_i\lambda_i'}\,,
\end{equation}
where $N_p$ is a normalization factor to ensure that the trace of the spin density matrix is one. The decay channel is then selected according to the branching ratios and the new particle momenta are generated according to
\begin{equation}
    \rho_{\lambda_0\lambda_0'}\mathcal{M}_{\lambda_0;\lambda_1\ldots\lambda_k}\mathcal{M}^*_{\lambda_0';\lambda_1'\ldots\lambda_k'}\prod_{i=1,k}D_{\lambda_i\lambda_i'}^i\,,
\end{equation}
where $\lambda_0$ is the helicity of the decaying particle and $\lambda_i$ is the helicity of the decay products. If there are any unstable particles in the above decay, they are selected as before and a spin density matrix is calculated and the process is repeated until only stable particles remain in the given chain. At this point, the decay matrix is calculated as
\begin{equation}
    D_{\lambda_0\lambda_0'}=\frac{1}{N_D}\mathcal{M}_{\lambda_0;\lambda_1\ldots\lambda_k}\mathcal{M}^*_{\lambda_0';\lambda_1'\ldots\lambda_k'}\prod_{i=1,n}D^i_{\lambda_i\lambda_i'}\,,
\end{equation}
where $N_D$ is chosen such that the trace of the decay matrix is one. Then another unstable particle is selected from the original decay and the process is repeated until the first decay chain ends in only stable particles. At this point, the next unstable particle is selected in the hard process and the above procedure repeats. Once there are only stable particles left, the procedure terminates.

\subsection{Achilles--Sherpa Interface}
Employing a dedicated version of the general-purpose event generator
Sherpa~\cite{Gleisberg:2003xi,Gleisberg:2008ta,Sherpa:2019gpd},
we construct an interface to the Comix matrix element
generator~\cite{Gleisberg:2008fv} to extract the leptonic current.
This interface has been described in detail in Ref.~\cite{Isaacson:2021xty}.
In order to provide the hard scattering amplitudes,
$\mathcal{M}_{\kappa_1\kappa_2;\lambda_1\ldots\lambda_n}$,
needed for the spin correlation algorithm in Sec.~\ref{sec:SpinDecays},
we make use of the methods developed in Ref.~\cite{Hoche:2014kca}.
This allows us to extract a spin-dependent leptonic current from Comix,
which can be contracted with the hadronic current obtained from Achilles.
Schematically this can be written as
\begin{equation}
  \mathcal{M}_{\kappa_h\kappa_\nu;\lambda_h\lambda_l\ldots\lambda_n}=
    g_{\mu\nu}\sum_{i} L^{(i)\,\mu}_{\kappa_\nu;\lambda_l\ldots\lambda_n}
    W^{(i)\,\nu}_{\kappa_h;\lambda_h}\;,
\end{equation}
where we have extended the notation of Ref.~\cite{Isaacson:2021xty}
to include spin labels.
As the spin states of the initial- and final-state hadrons are not
observed experimentally, they can be averaged and summed over,
leading to the final expression
\begin{equation}
  \mathcal{M}_{\kappa_\nu;\lambda_l\ldots\lambda_n}
  \mathcal{M}_{\kappa_\nu';\lambda_l'\ldots\lambda_n'}^*=
    \frac{1}{2}\,g_{\mu\nu}g_{\mu'\nu'}\sum_{i,i'}
    L^{(i)\,\mu}_{\kappa_\nu;\lambda_l\ldots\lambda_n}
    L^{(i')\,\mu'}_{\kappa_\nu';\lambda_l'\ldots\lambda_n'}
    W^{(i)\,\nu}_{\kappa_h;\lambda_h}
    W^{(i')\,\nu'}_{\kappa_h';\lambda_h'}
    \delta_{\kappa_h\kappa_h'}\,
    \delta_{\lambda_h\lambda_h'}\;.
\end{equation}
The resulting tensor is inserted into the event record of Sherpa
and used to seed the event generation algorithms described in
Ref.~\cite{Hoche:2014kca,Laubrich:2006aa}, which accounts for
all spin correlations along all decay chains. We note that this procedure
is independent of the physics model for the short-distance interactions,
and that arbitrary beyond Standard Model scenarios can easily be 
implemented by providing the corresponding UFO output~\cite{Degrande:2011ua}
of FeynRules~\cite{Christensen:2008py,Alloul:2013bka}.

\section{Results}
\label{sec:Results}

We consider the scattering of a tau neutrino off an argon nucleus through the use of a rescaled
carbon spectral function for both a monochromatic beam (for validation) and for a realistic flux at DUNE. 
For this study, we focus only on the quasielastic region for the nuclear interaction, as implemented in
Ref.~\cite{Isaacson:2022cwh}, and we neglect
final state interactions. Final state interactions will modify the 2 and 3 pion distributions and investigating
the size of the changes is left to a future work. For reference, all tau lepton decay channels with a branching ratio 
above 0.5\% are given in Tab.~\ref{tab:decays}. However, all possible decays are actually included in our simulation.
\begin{table}[ht]
    \centering
    \begin{tabular}{|c|c|}
    \hline
    Decay mode     &  Branching ratio (\%)\\
    \hline
    \hline
    Leptonic decays & 35.21 \\
    \hline
    $e^-\nu_\tau\bar{\nu}_e$     & 17.85 \\
    $\mu^-\nu_\tau\bar{\nu}_\mu$     & 17.36 \\
    \hline
    \hline
    Hadronic decays & 64.79 \\
    \hline
    $\pi^-\pi^0\nu_\tau$ & 25.50 \\
    $\pi^-\nu_\tau$ & 10.90 \\
    $\pi^+\pi^-\pi^-\nu_\tau$ & 9.32 \\
    $\pi^-\pi^0\pi^0\nu_\tau$ & 9.17 \\
    $\pi^+\pi^-\pi^-\pi^0\nu_\tau$ & 4.50 \\
    $\pi^-\pi^0\pi^0\pi^0\nu_\tau$ & 1.04 \\
    $K^- \nu_\tau$ & 0.70 \\
    $\pi^+\pi^-\pi^-\pi^0\pi^0$ & 0.55 \\
    other & 3.11 \\
    \hline
    \end{tabular}
    \caption{Decay channels of the tau lepton with branching fractions greater than 0.5\%.
      All other channels are grouped into the ``other'' category.}
    \label{tab:decays}
\end{table}

The spectral function used in this calculation was obtained within the correlated basis function theory 
of Ref.~\cite{Benhar:1994hw}. Electron scattering data is used to constrain the low momentum and
energy contributions in the mean-field calculations. The correlated component is obtained
within the Local Density Approximation. The normalization of the spectral function is taken as
\begin{equation}
    \int \frac{{\rm d}k_h}{(2\pi)^3} {\rm d}E S_h(\vec{k}_h, E) = 
    \begin{cases}
    Z, & h = p\,, \\
    A-Z, & h=n\,,
    \end{cases}
\end{equation}
where $k_h$ is the momentum of the initial nucleon, $E$ is the removal energy, $S_h$ is the spectral function,
and $Z (A)$ denotes the number of protons (nucleons) in the nucleus.

In this work, we consider the Kelly parametrization for the electric and magnetic form factors~\cite{Kelly:2004hm},
and use a dipole axial form factor with $g_A=1.2694$ and $M_A=1.0$ GeV. Additionally, the pseudoscalar
form factor is obtained through the use of the partially conserved axial current ansatz and assumptions about
the pion-pole dominance, \textit{i.e.}
\begin{equation}
    F^A_P(Q^2) = \frac{2 m_N^2}{Q^2 + m_\pi^2} F^A(Q^2)\,,
\end{equation}
where $F^A_P$ is the pseudoscalar axial form factor, $m_N, m_\pi$ are the masses of the nucleon and pion, 
respectively, $Q^2 = -q^2$ is the momentum transfer, and $F^A$ is the axial form factor.

\subsection{Monochromatic beam}\label{subsec:mono}

\begin{figure}
    \centering
    \includegraphics[width=0.48\textwidth]{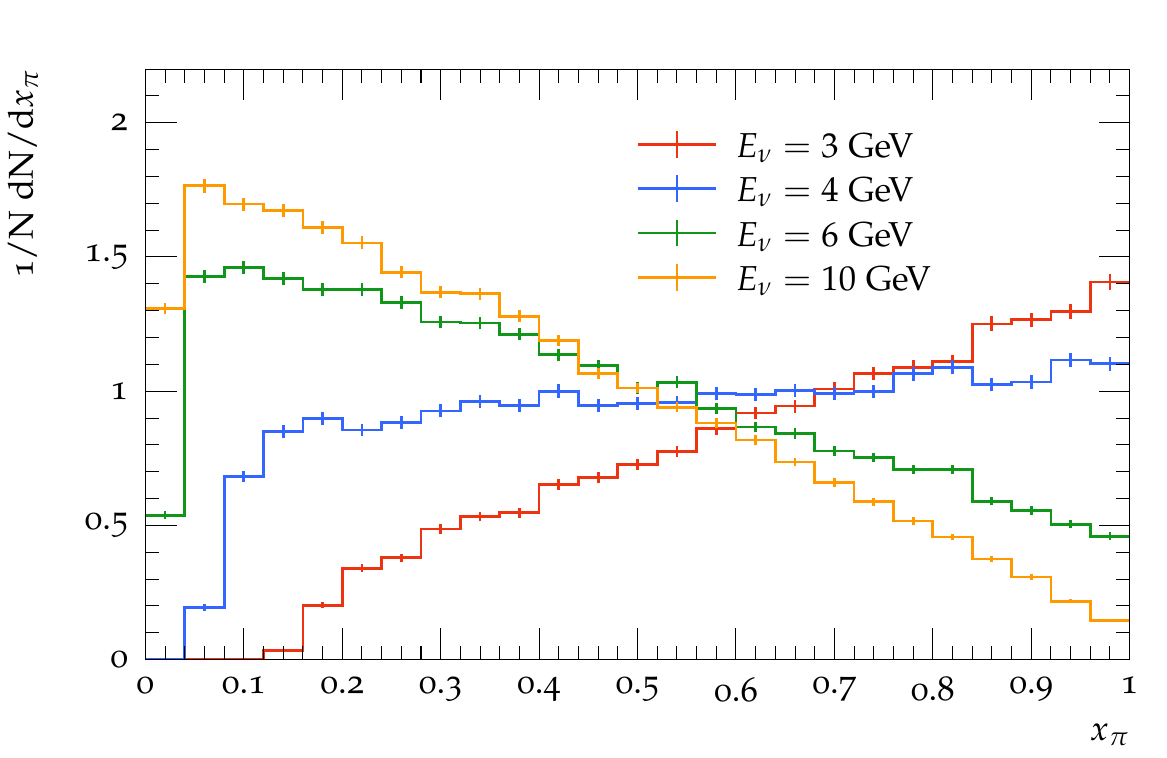}
    \hfill
    \includegraphics[width=0.48\textwidth]{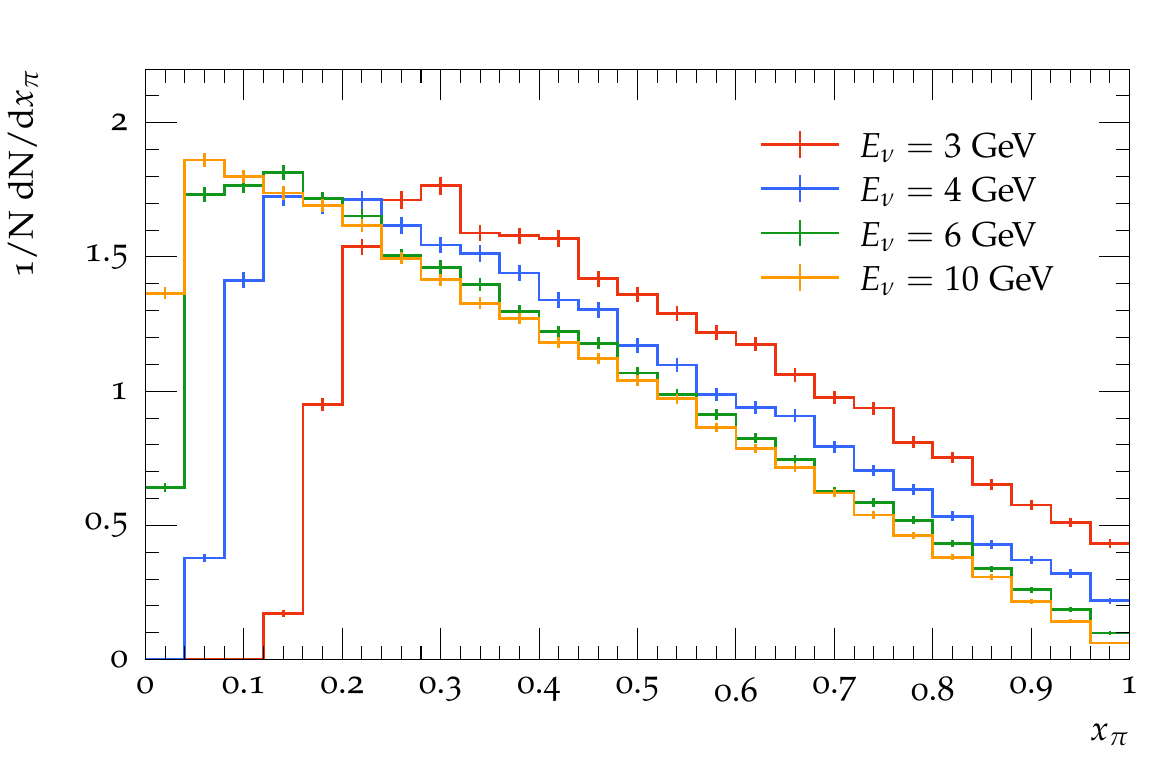}
    \caption{Momentum fraction of the outgoing pion for $\tau^{-} \rightarrow \pi^{-} \nu_\tau$ decays of 
    various incoming neutrino energies. Results are shown for the full polarization calculation on the left and the left-handed polarization approximation ($P_{L}^{T}=1, P_{T}^{T}=0$) on the right. }
    \label{fig:xpi}
\end{figure}

In order to validate our results, we first consider monochromatic beams. We compare our calculations 
to the results from Ref.~\cite{Hernandez:2022nmp} for the single pion production channel.
However, instead of the momentum of the outgoing pion, we analyze the momentum fraction
of the outgoing pion ($x_\pi = p_\pi / p_\tau$). This allows us to include multiple neutrino energies in
the same plot. The results from Achilles+Sherpa are shown in Fig.~\ref{fig:xpi}, with the appropriate handling
of the tau polarization on the left and assuming the tau to be purely left-handed on the right. From this,
we see that our results are consistent with those from Ref.~\cite{Hernandez:2022nmp}. Additionally, we see that
as the neutrino energy increases the results approach those found in Fig.~\ref{fig:xpi_collinear} for
the collinear limit, as expected.

We next consider the decays of the tau into the two pion and three pion states, which are dominated by
the decay chain $\tau^- \to \nu_\tau \rho^- (\rho^- \to \pi^- \pi^0)$ and
$\tau^- \to \nu_\tau a_1^- (a_1^- \to \pi^- \pi^- \pi^+$ or $a_1^- \to \pi^- \pi^0 \pi^0)$ respectively.

\begin{figure}
    \centering
    \includegraphics[width=0.48\textwidth]{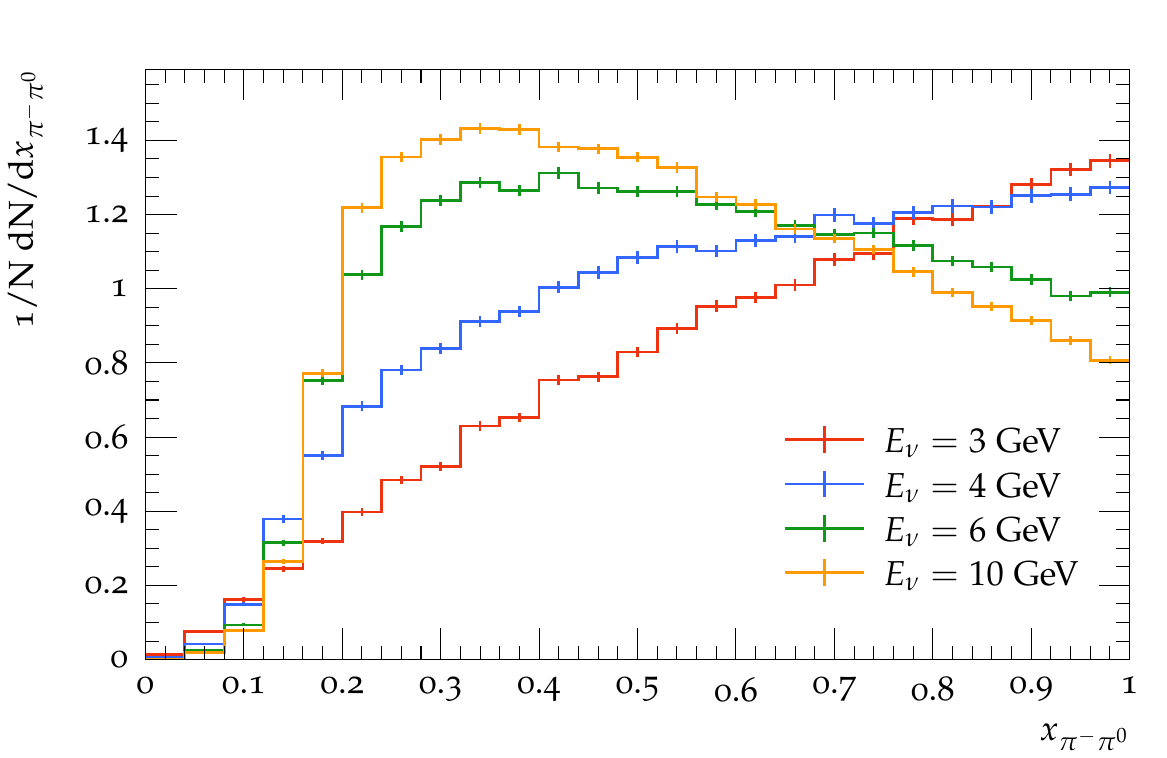}
    \hfill
    \includegraphics[width=0.48\textwidth]{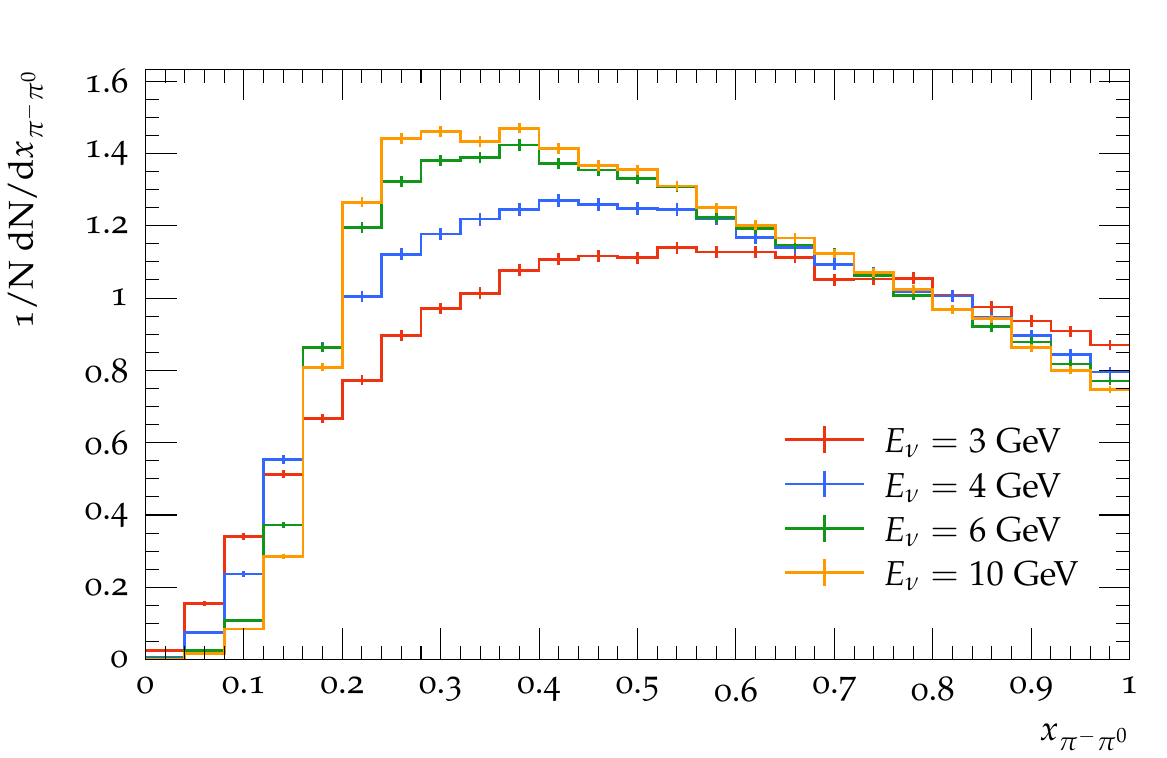}
    \caption{Momentum fraction of the $\pi^{-}\pi^{0}$ system for $\tau^{-} \rightarrow \pi^{-} \pi^{0} \nu_\tau$ decays of 
    various incoming neutrino energies. Results are shown for the full polarization calculation on the left and the left-handed polarization approximation ($P_{L}^{T}=1, P_{T}^{T}=0$) on the right. }
    \label{fig:xpipi}
\end{figure}

\begin{figure}  
    \centering
    \includegraphics[width=0.48\textwidth]{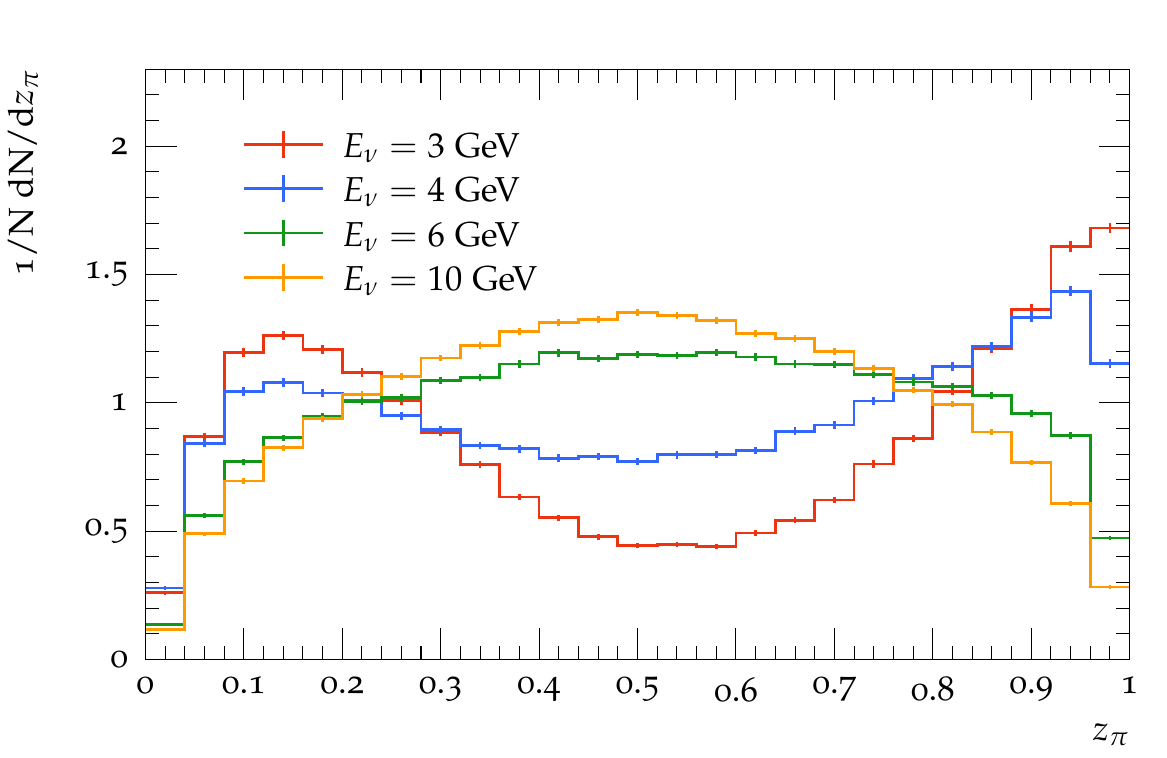}
    \hfill
    \includegraphics[width=0.48\textwidth]{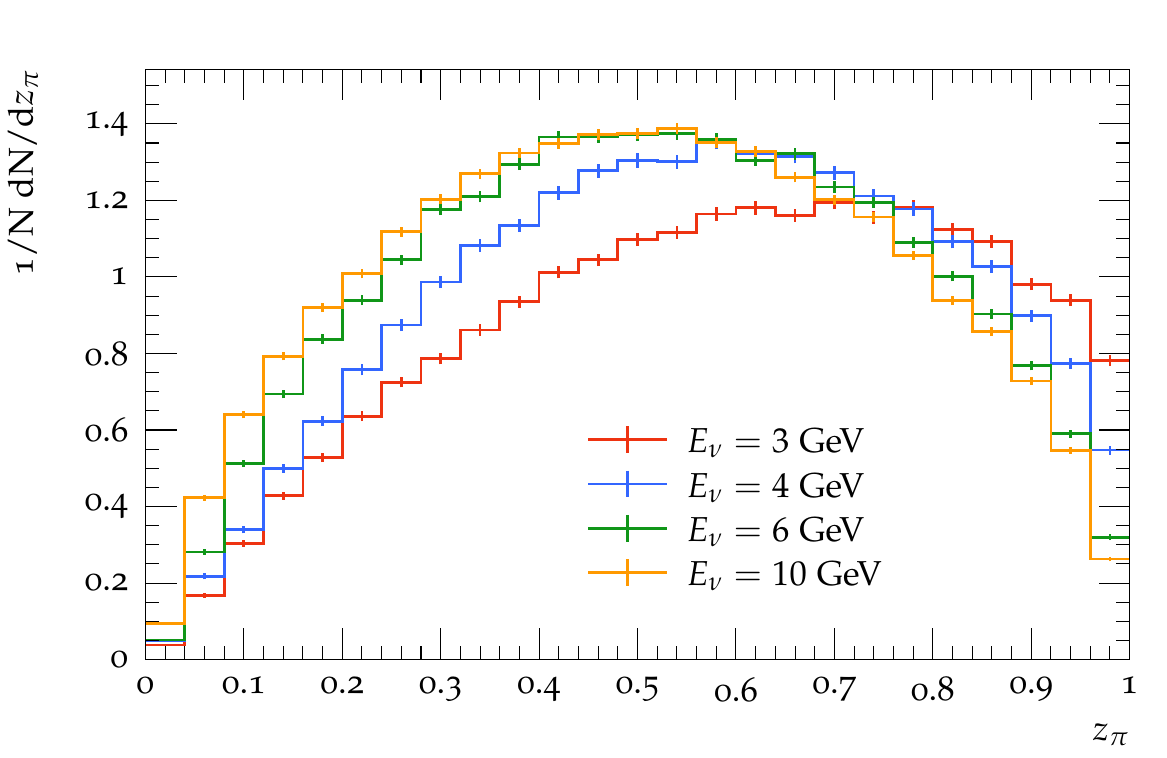}
    \caption{Ratio of the $\pi^{-}$ momentum to the $\rho^{-}$ momentum for $\tau^{-} \rightarrow \pi^{-} \pi^{0} \nu_\tau$ decays of 
    various incoming neutrino energies, where $z_{\pi}$ denotes this ratio. Results are shown for the full polarization calculation on the left and the left-handed polarization approximation ($P_{L}^{T}=1, P_{T}^{T}=0$) on the right. }
    \label{fig:zpi}
\end{figure}

For the case of the $\rho$ channel, we analyze the momentum fraction of the hadronic system
($x_\rho = p_\rho/p_\tau$) as well as the momentum fraction of the $\pi^-$ with respect to the $\rho$
($z_{\pi} = p_{\pi^-} / p_\rho$). The results are shown in Fig.~\ref{fig:xpipi} and Fig.~\ref{fig:zpi}
respectively. Again, the full calculation is on the left of each plot and the assumption of a purely left-handed
tau is on the right. We can see that there is a significant impact from including the correct polarization
in the calculation. In the case of the $\rho$ momentum fraction, we see that our results approach the
transverse curve for the $\rho$ from Fig.~\ref{fig:xpi_collinear} as $E_\nu$ increases.
This is expected since we are summing over the polarizations of the $\rho$, which
are dominated by the transverse polarization.

\begin{figure}  
    \centering
    \includegraphics[width=0.48\textwidth]{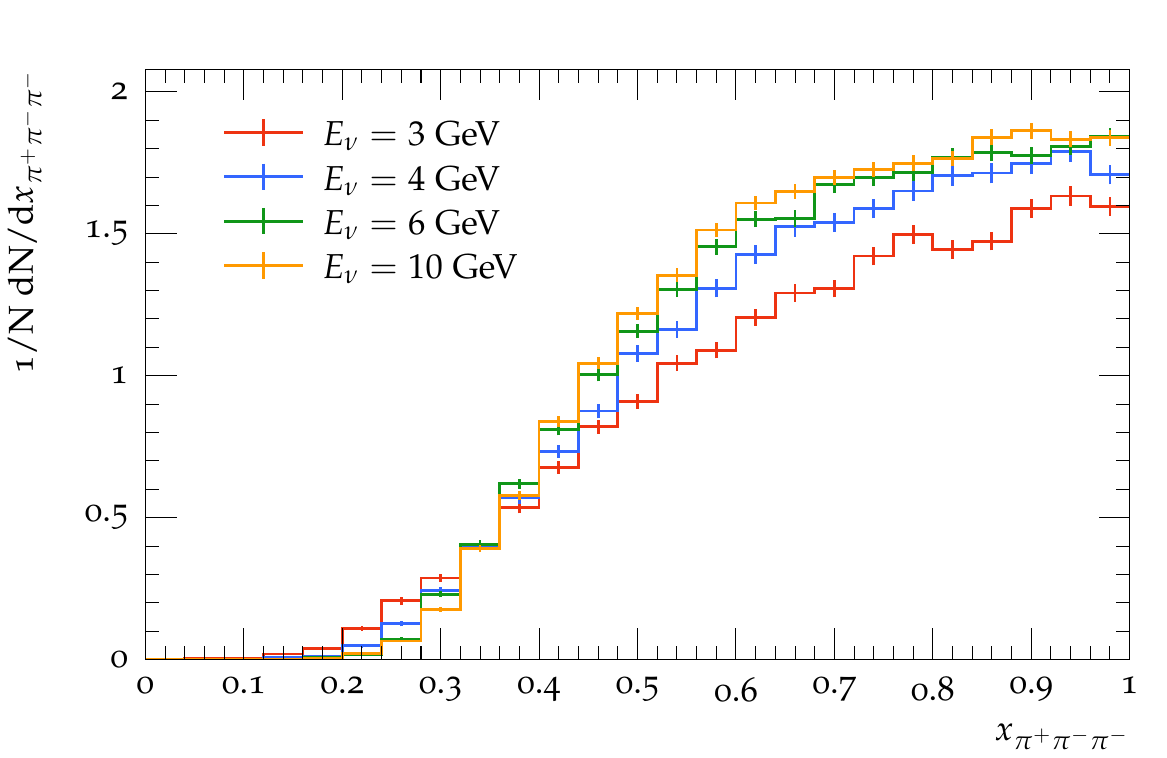}
    \hfill
    \includegraphics[width=0.48\textwidth]{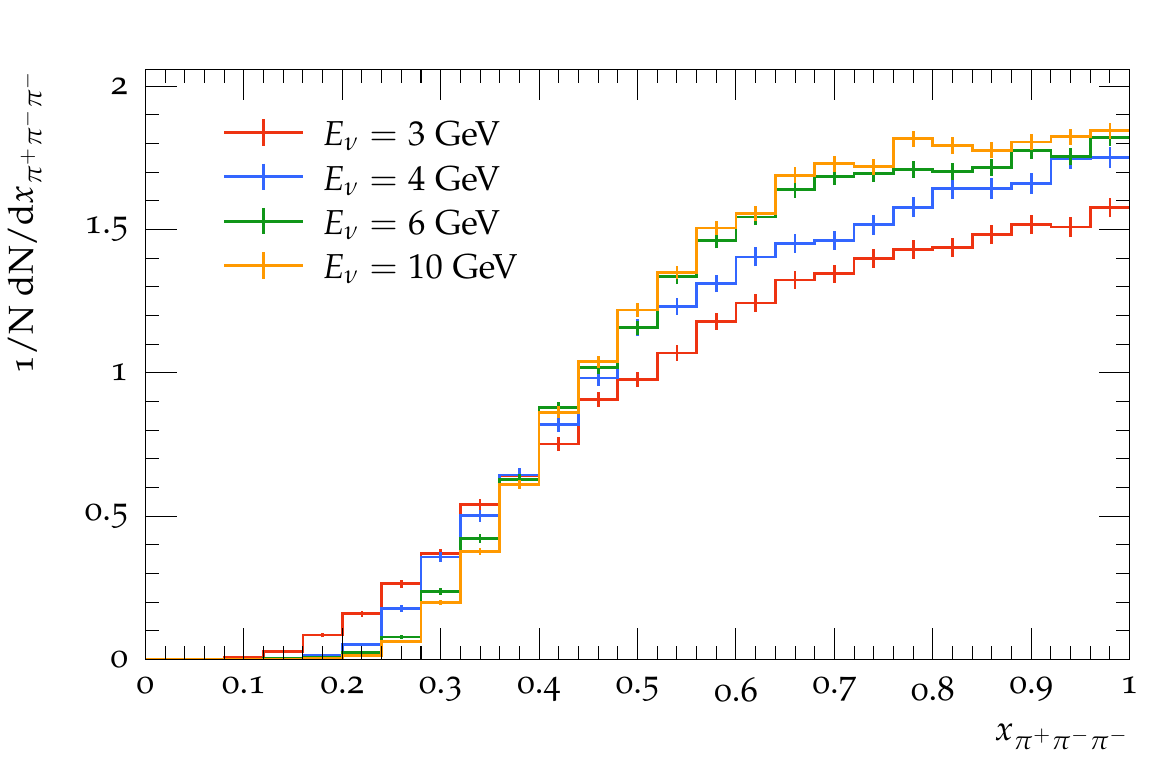}
    \caption{Momentum fraction of the $\pi^{+}\pi^{-}\pi^{-}$ system for $\tau^{-} \rightarrow \pi^{+} \pi^{-} \pi^{-} \nu_\tau$ decays of 
    various incoming neutrino energies. Results are shown for the full polarization calculation on the left and the left-handed polarization approximation ($P_{L}^{T}=1, P_{T}^{T}=0$) on the right. }
    \label{fig:xa1_pip}
\end{figure}

\begin{figure}  
    \centering
    \includegraphics[width=0.48\textwidth]{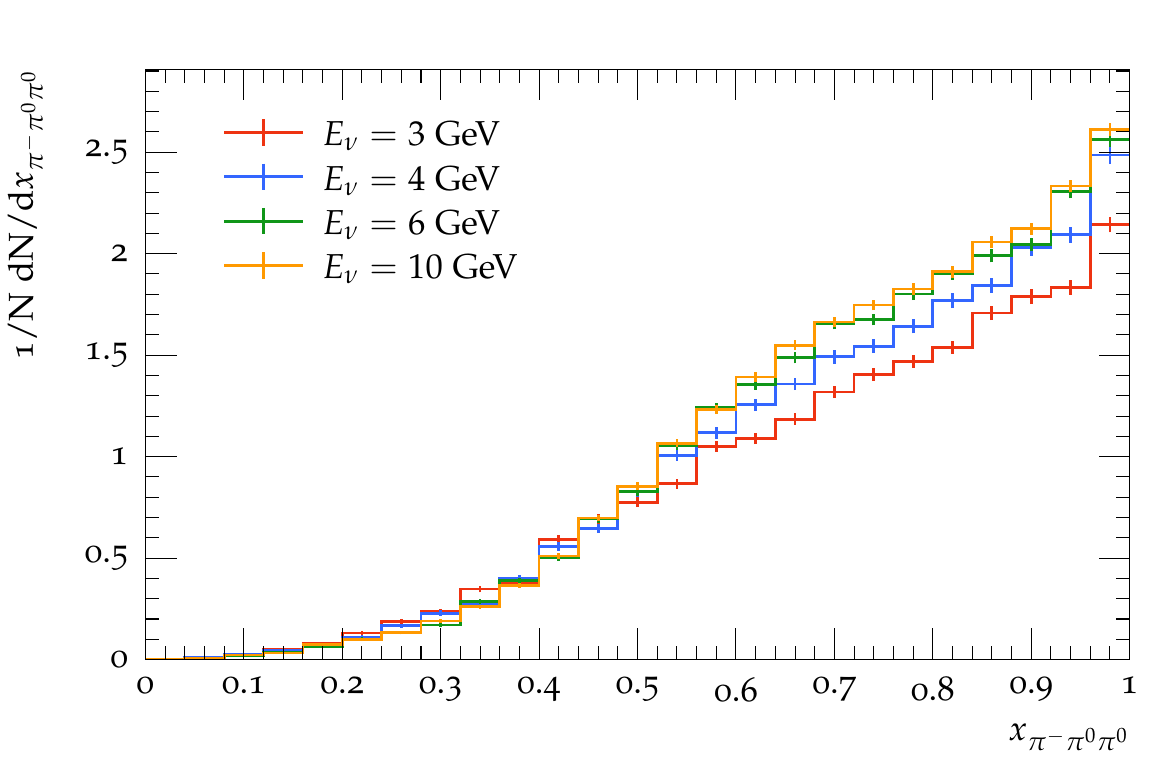}
    \hfill
    \includegraphics[width=0.48\textwidth]{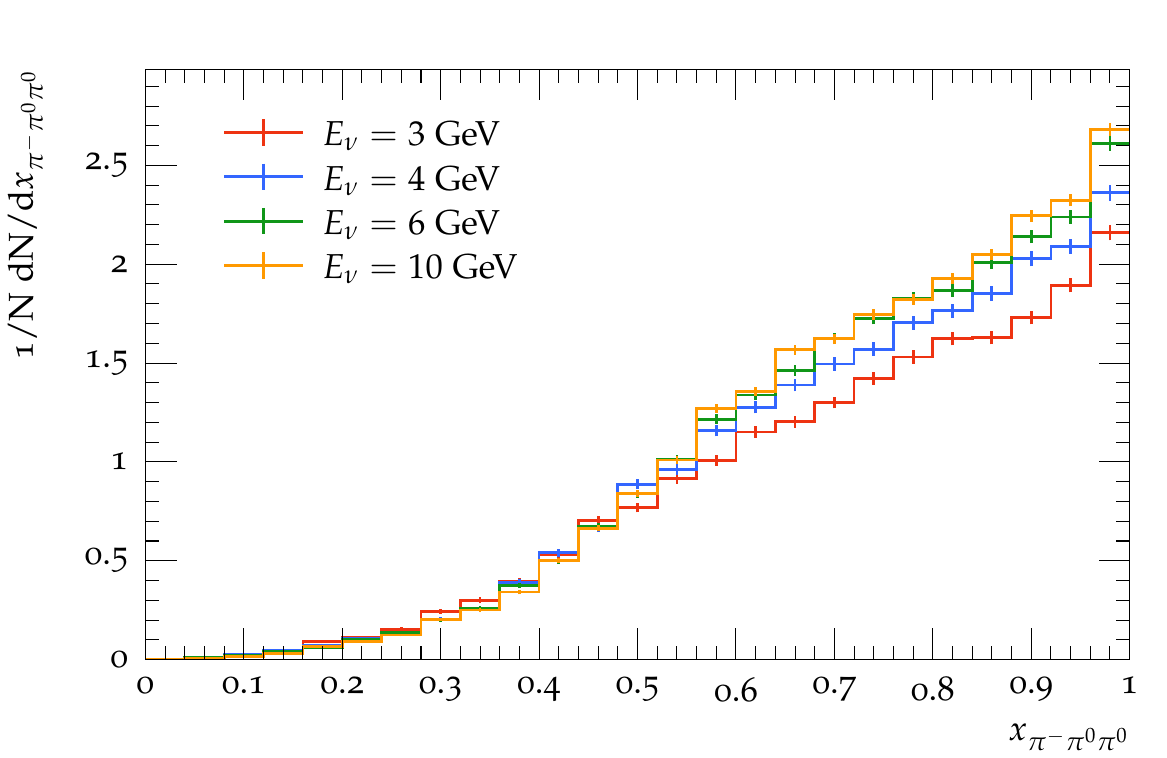}
    \caption{Momentum fraction of the $\pi^{-}\pi^{0}\pi^{0}$ system for $\tau^{-} \rightarrow \pi^{-} \pi^{0} \pi^{0} \nu_\tau$ decays of 
    various incoming neutrino energies. Results are shown for the full polarization calculation on the left and the left-handed polarization approximation ($P_{L}^{T}=1, P_{T}^{T}=0$) on the right. }
    \label{fig:xa1_pim}
\end{figure}

As mentioned in Sec.~\ref{sec:collinear_limit}, summing over the polarizations of the $a_1$ removes any sensitivity
to the polarization of the $\tau$. Therefore, the $a_1$ momentum as a fraction of the $\tau$ momentum ($x_{a_1} = p_{a_1} / p_\tau$)
should not show any difference between the full calculation and the left-handed only calculation.
This is supported by Figs.~\ref{fig:xa1_pip} and~\ref{fig:xa1_pim}, with the left and right panel
being statistically consistent with
each other. Figure~\ref{fig:xa1_pip} shows the decay to the $\pi^+\pi^-\pi^-$ final state and Fig.~\ref{fig:xa1_pim}
shows the decay to the $\pi^-\pi^0\pi^0$ final state.
Furthermore, the curves approach the result of the collinear limit as $E_\nu$ increases, as seen by comparing to
the transverse $a_1$ curve of Fig.~\ref{fig:xpi_collinear}.

\begin{figure}  
    \centering
    \includegraphics[width=0.48\textwidth]{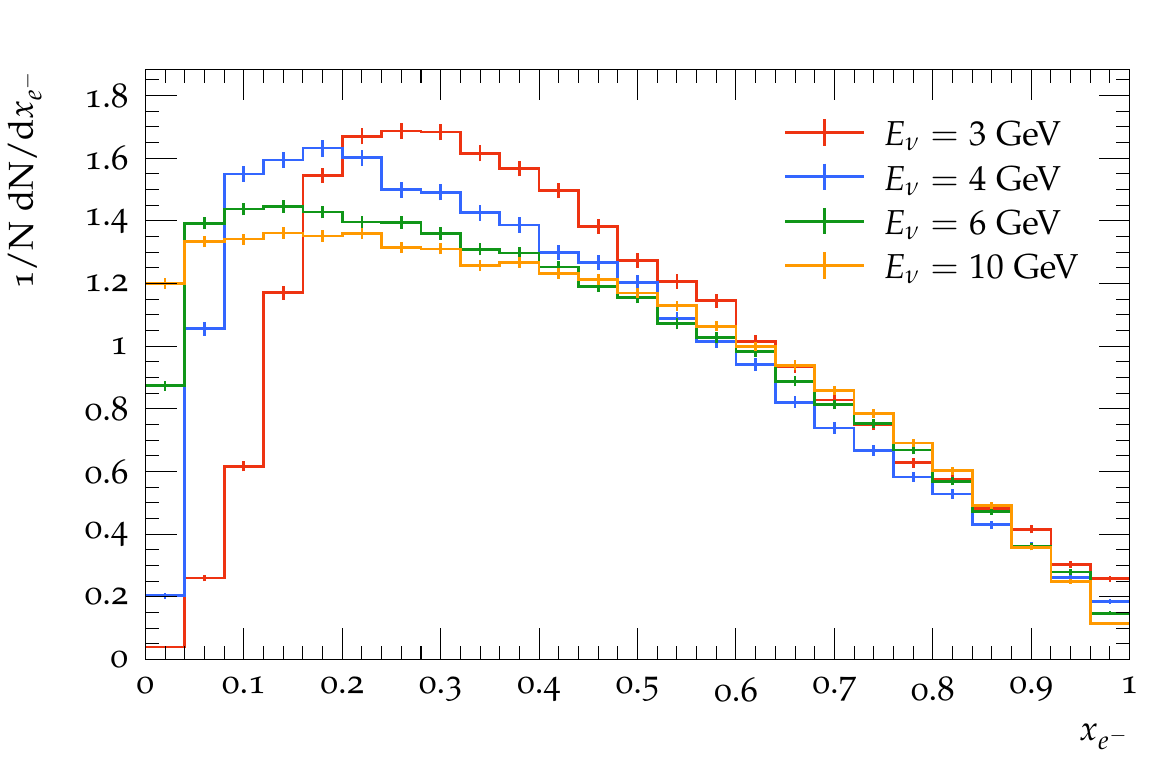}
    \hfill
    \includegraphics[width=0.48\textwidth]{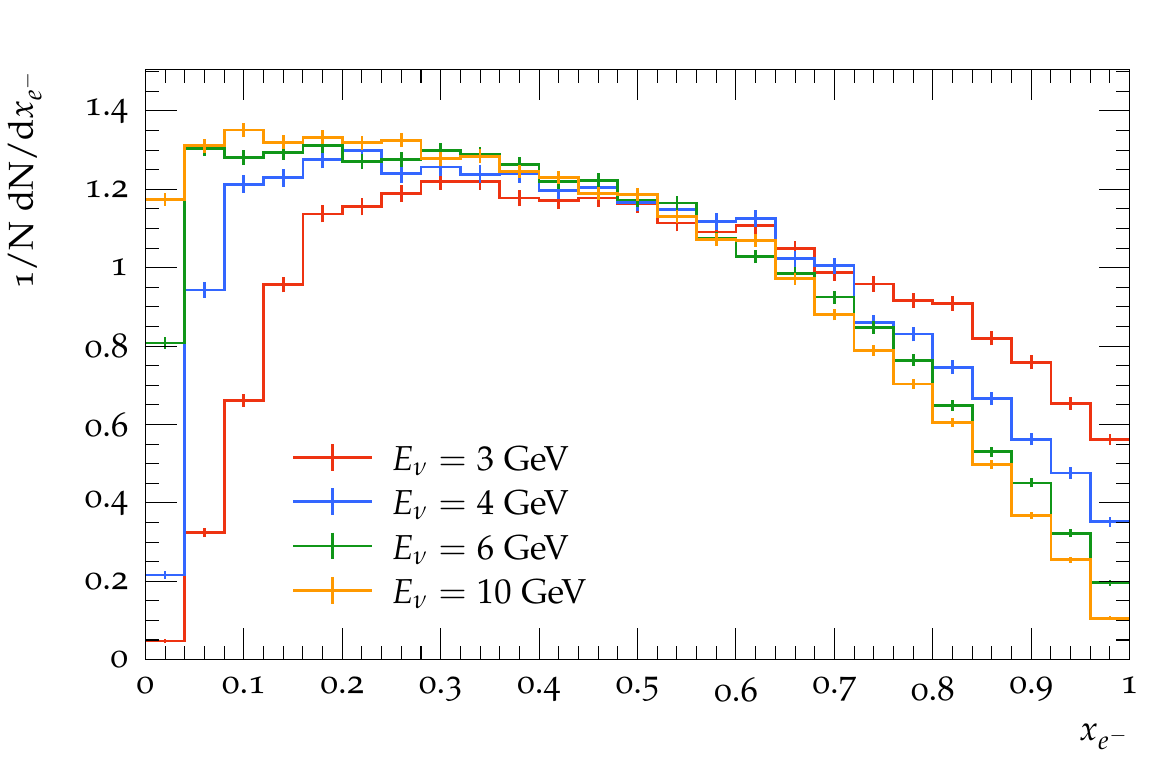}
    \caption{Momentum fraction of the outgoing electron for $\tau^{-} \rightarrow e^{-} \nu_\tau \bar{\nu}_e$ decays of 
    various incoming neutrino energies. Results are shown for the full polarization calculation on the left and the left-handed polarization approximation ($P_{L}^{T}=1, P_{T}^{T}=0$) on the right. }
    \label{fig:xlep}
\end{figure}

Finally, we consider the leptonic decay channel. Here we will focus on the decays to electrons due to the
possible experimental relevance at DUNE for $\nu_\tau$ detection, but note that up to corrections from the
muon mass and the difference in the branching ratios the predictions would be identical. The comparison
for various neutrino energies is given in Fig.~\ref{fig:xlep}. Again, we can see a difference between
the full calculation in the left panel and the purely left-handed calculation in the right panel.
The latter result approaches the expected prediction for large $E_\nu$ as shown in Fig.~\ref{fig:xpi_collinear}.

\subsection{Realistic beams}

\begin{figure}
    \centering
    \includegraphics[width=0.5\textwidth]{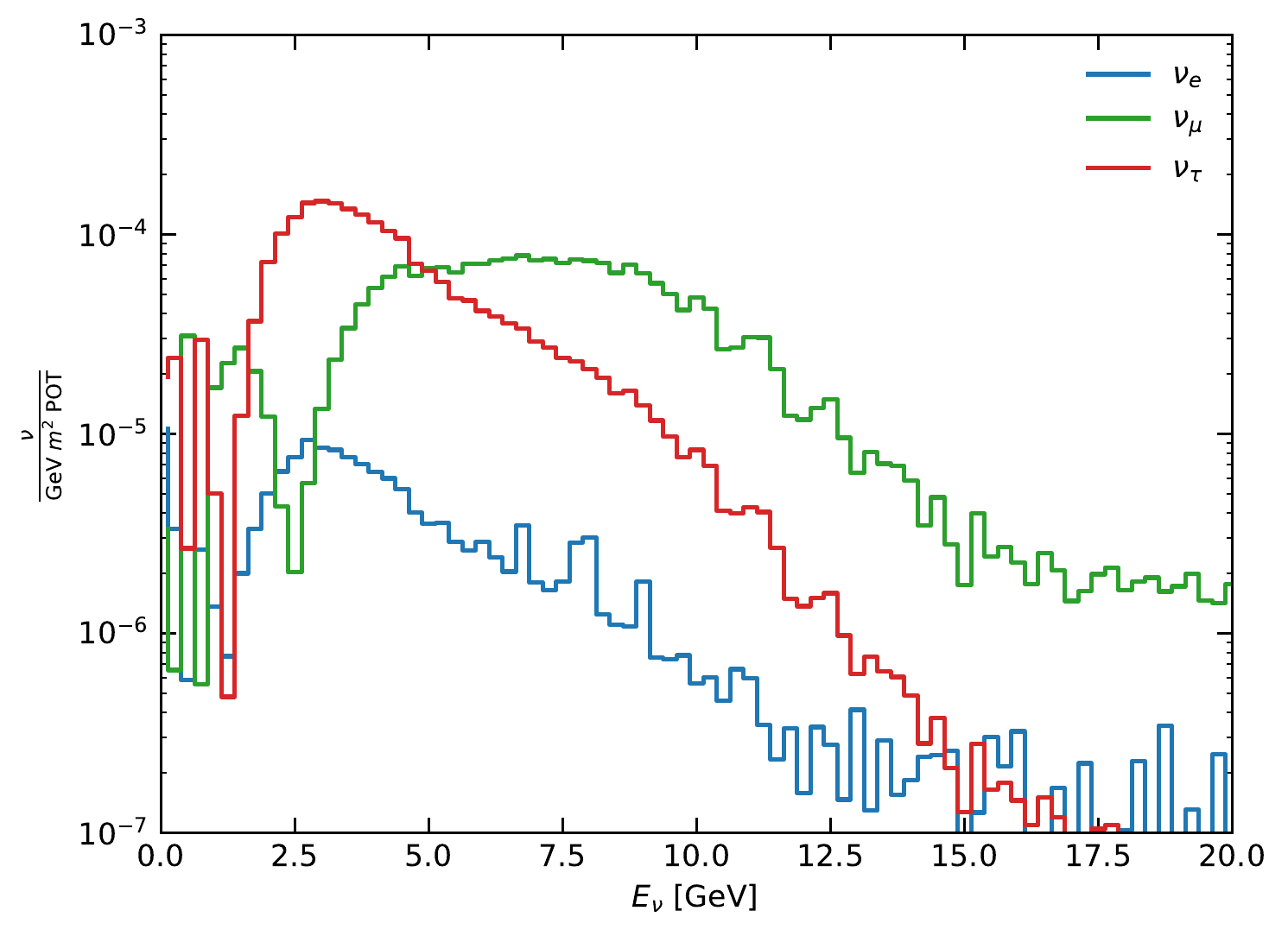}
    \caption{Neutrino flux in the far detector of DUNE. The flux is generated from running in 
    $\tau$-optimized mode. The unoscillated fluxes are obtained from Ref.~\cite{FieldsDune}.}
    \label{fig:flux}
\end{figure}

To investigate the impact of spin-correlations in a more realistic setting, we consider the $\tau$-optimized flux mode
for the DUNE experiment~\cite{FieldsDune}. The oscillated far detector flux is shown in Fig.~\ref{fig:flux}. The oscillation parameters
are fixed to the values from the global fit~\cite{deSalas:2020pgw}:
\begin{align*}
    \delta m_{21}^2 &= 7.50 \times 10^{-5}\ \textrm{eV}^2,\quad \delta m_{31}^2 = 2.55 \times 10^{-3} \textrm{eV}^2, \\
    s_{12}^2 &= 0.318,\ s_{23}^2 = 0.574,\ s_{13}^2 = 0.0220,\ \delta_{CP} = 1.08\pi\,.
\end{align*}
The results are given using the flux averaged cross-section, defined as
\begin{equation}
    \langle\sigma\rangle = \frac{\int {\rm d}E_\nu \Phi(E_\nu) \sigma(E_\nu)}
                                {\int {\rm d}E_\nu \Phi(E_\nu)}\,,
\end{equation}
where $\Phi(E_\nu)$ is the neutrino flux and $\sigma(E_\nu)$ is the neutrino energy dependent cross-
section. 

While all possible decay channels are implemented, we consider here only those most affected by 
correctly handling polarization. Furthermore, only decay channels with sufficiently large branching ratios
such that the differences are experimentally relevant are shown.

\begin{figure}
    \centering
    \includegraphics[width=0.48\textwidth]{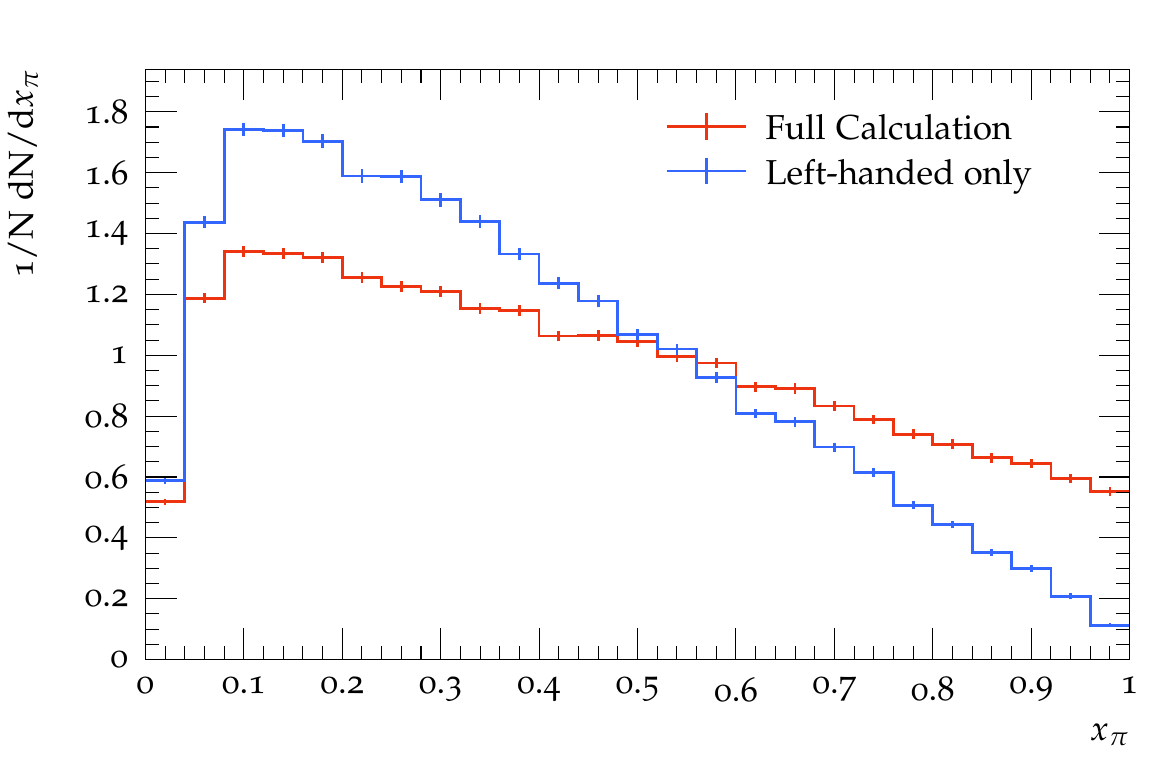} \hfill
    \includegraphics[width=0.48\textwidth]{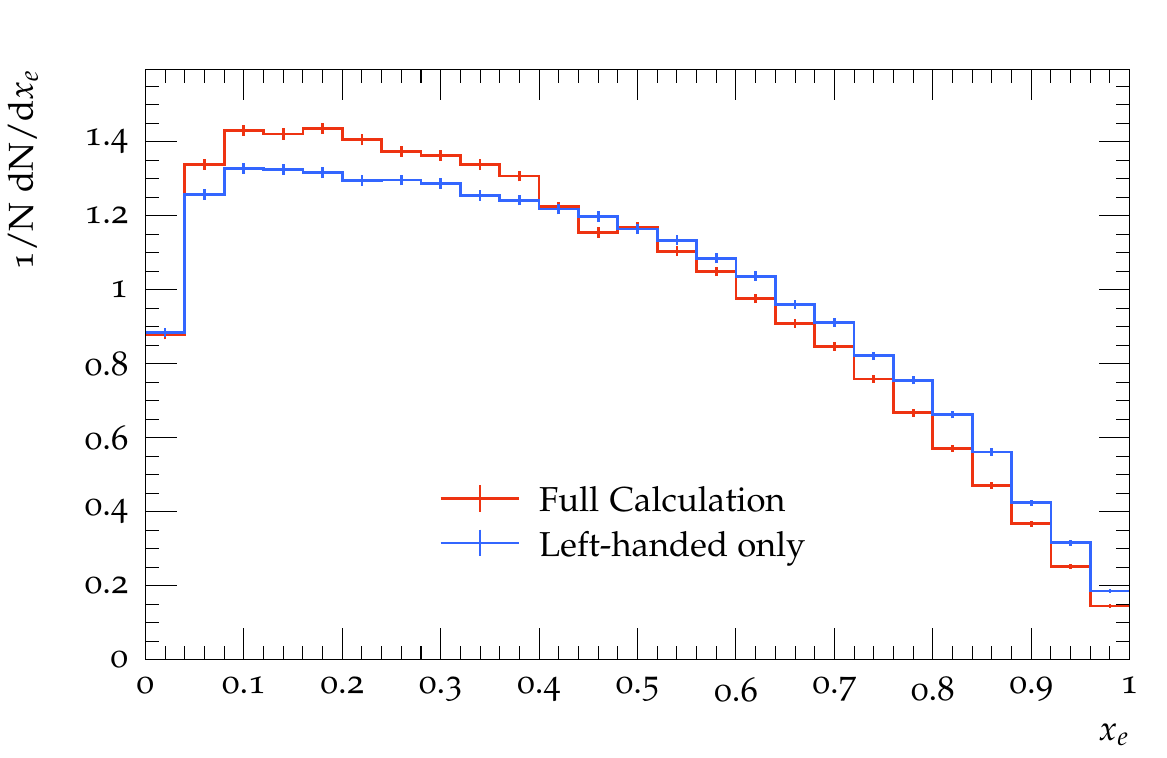} 
    \caption{Momentum fraction distribution for the decay of the $\tau$ into a single pion is shown on the 
    left and momentum fraction distribution for the decay into an electron is shown on the right.
    The full polarization handling is shown in red with the approximation that
    the $\tau$ is purely left-handed in blue. The predictions are folded over the DUNE far-detector flux
    running in the $\tau$-optimized mode given in Fig.~\ref{fig:flux}.}
    \label{fig:dune_xpi_xenu}
\end{figure}

We first consider the single pion decay channel, since it is a clean channel to reconstruct at DUNE.
The results of the calculation are shown in the left panel of Fig.~\ref{fig:dune_xpi_xenu}.
Here we see that in the full calculation, the outgoing pion tends to be more energetic than in the
fully left-handed case.

The case of leptonic decays is shown in the right panel of Fig.~\ref{fig:dune_xpi_xenu}, and is
calculated in the massless limit for both the electron and the muon. In this case, the two decays
are identical. The effect of including the full polarization information makes the outgoing lepton
softer compared to the fully left-handed calculation.
While the chance of detecting the muon channel is extremely difficult, there is
a chance to detect the electron channel due to the low $\nu_e$ flux at the far detector as seen in
Fig.~\ref{fig:flux}.

\begin{figure}
    \centering
    \includegraphics[width=0.48\textwidth]{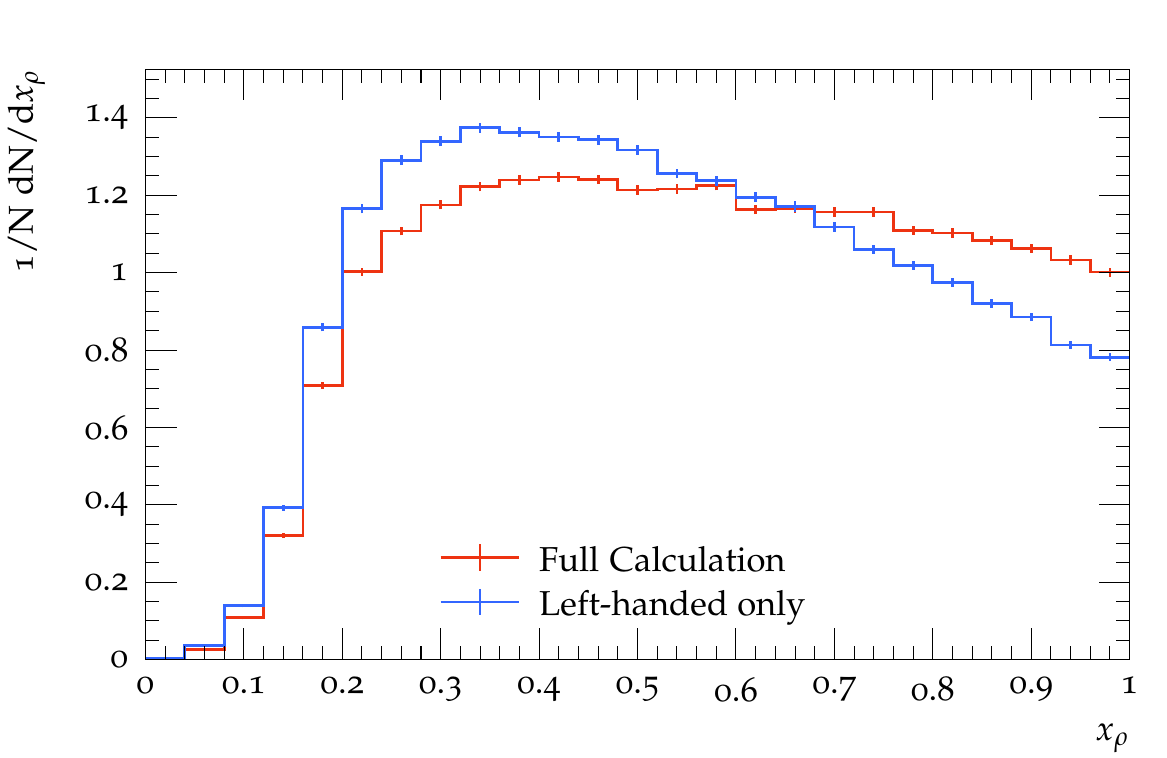} \hfill
    \includegraphics[width=0.48\textwidth]{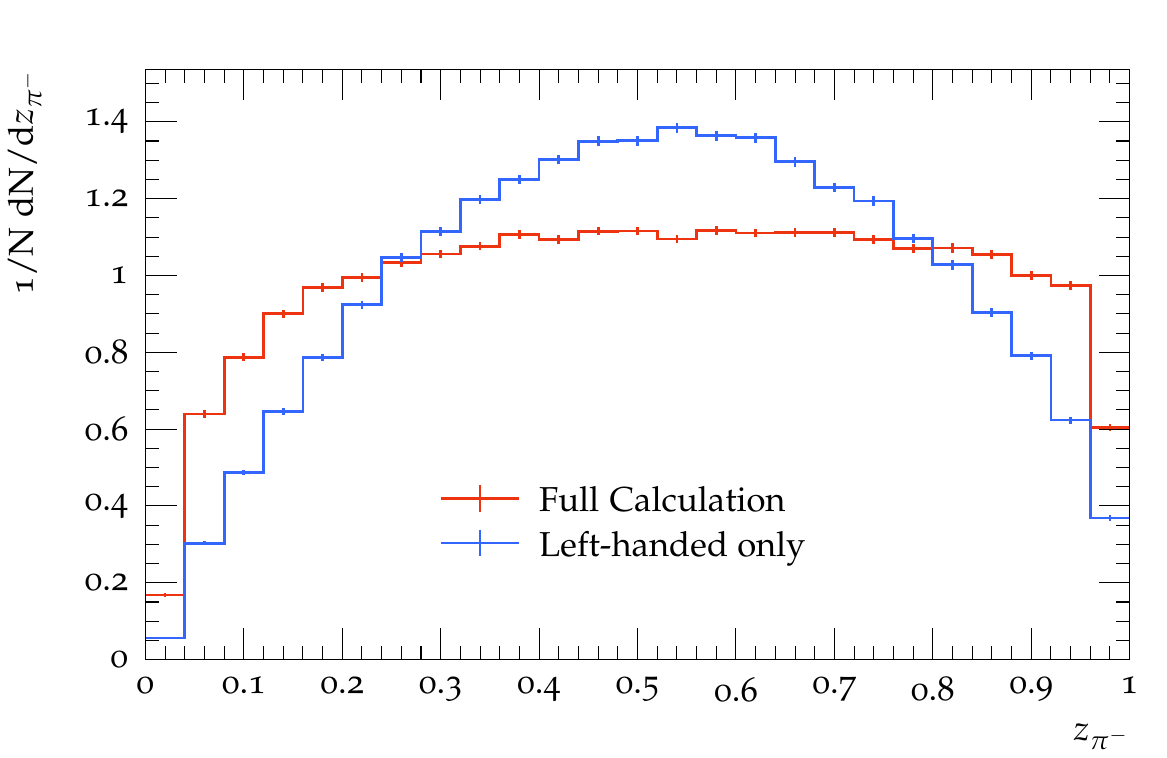} 
    \caption{Momentum fraction distribution for the decay of the $\tau$ into a pair of pions is shown
    on the left. The momentum of the negatively charged pion as a fraction of the sum of the pion momenta
    is given on the right.
    The full polarization handling is shown in red with the approximation that
    the $\tau$ is purely left-handed in blue. The predictions are folded over the DUNE far-detector flux
    running in the $\tau$-optimized mode given in Fig.~\ref{fig:flux}.}
    \label{fig:dune_rho}
\end{figure}

Another interesting decay channel to consider is the two pion final state, which has the largest
branching fraction of all decay channels. For this decay channel, we consider
the momentum of the sum of the two pions as a fraction of the $\tau$ momentum ($x_{\pi\pi}$)
and the momentum of the negatively charged pion as a fraction of the momentum sum ($z_{\pi}$).
Figure~\ref{fig:dune_rho} shows the difference between the full calculation in red and the fully
left-handed approximation in blue. In the case of the $x_{\pi\pi}$ distribution, the total momentum
is harder in the full calculation compared to the left-handed assumption. Additionally, there is a
significant difference in $z_{\pi}$ between the full calculation and the left-handed only. The 
full calculation is relatively flat over the full range, while the left-handed only calculation is
peaked around 0.6. This shift is significant, and will be important for any detailed study on using
the two pion channel to detect tau neutrino events.

\begin{figure}
    \centering
    \includegraphics[width=0.48\textwidth]{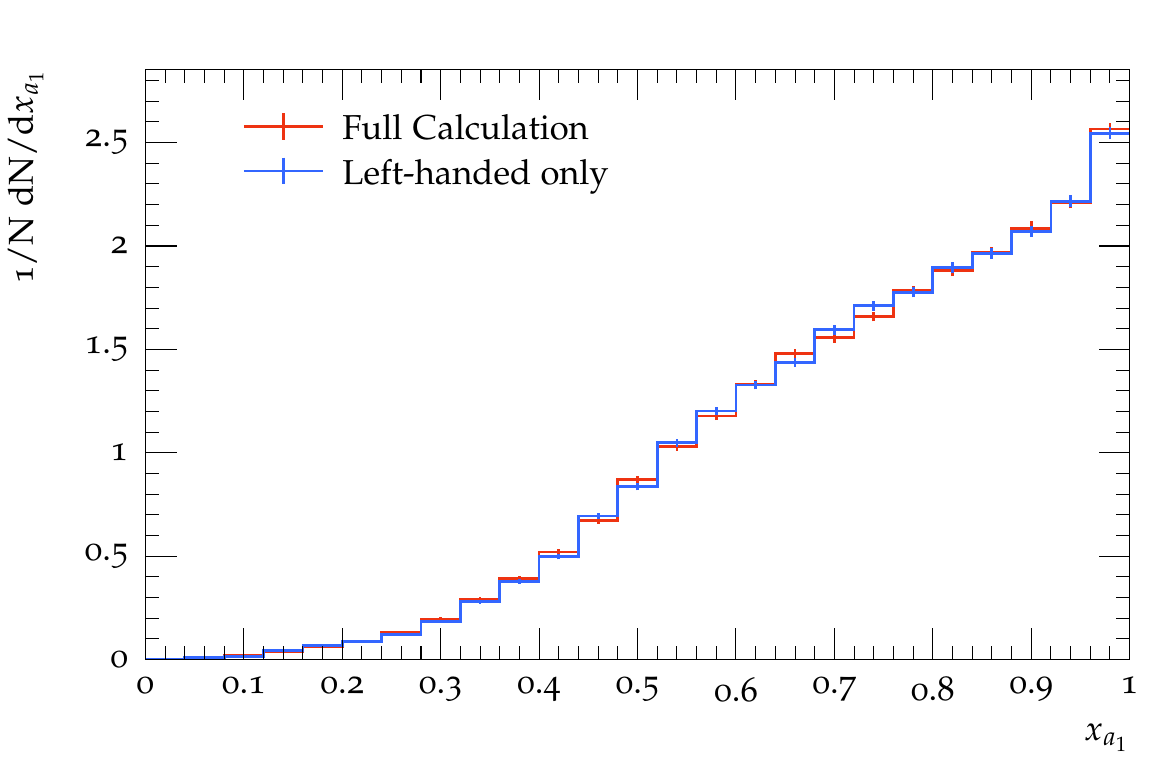} \hfill
    \includegraphics[width=0.48\textwidth]{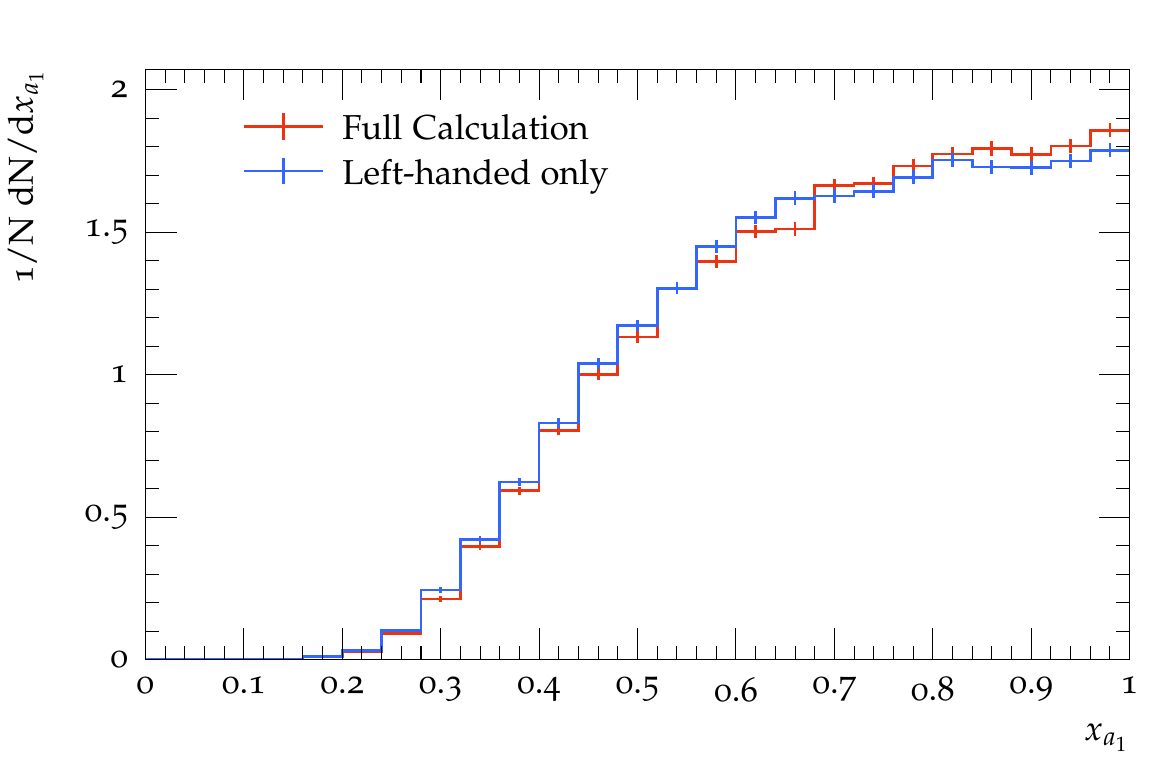}
    \caption{The full calculation (red) and the purely left-handed calculation (blue) are given for the 
    momentum fraction of the three pions as a fraction of the total $\tau$
    momentum for the decay of the $a_1$, with the $\pi^0\pi^0\pi^-$ channel on the left 
    and the $\pi^+\pi^-\pi^-$ channel on the right.  The predictions are folded over the DUNE far-detector flux
    running in the $\tau$-optimized mode given in Fig.~\ref{fig:flux}.}
    \label{fig:dune_a1}
\end{figure}

The last decay channel considered in this work is the decay to three pions. In this case, the decay is dominated
by the $a_1$ meson as discussed in Sec.~\ref{sec:collinear_limit}, and since we are not separating out the $a_1$
polarization should not be sensitive to the polarization of the $\tau$. This can be seen in Fig.~\ref{fig:dune_a1},
where the decay $a_1 \to \pi^0\pi^0\pi^-$ can be seen on the left and the decay $a_1 \to \pi^+\pi^-\pi^-$ can be seen
on the right. The full calculation and the left-handed only calculation are statistically consistent with each
other, as expected.

\begin{figure}
    \centering
    \includegraphics[width=0.48\textwidth]{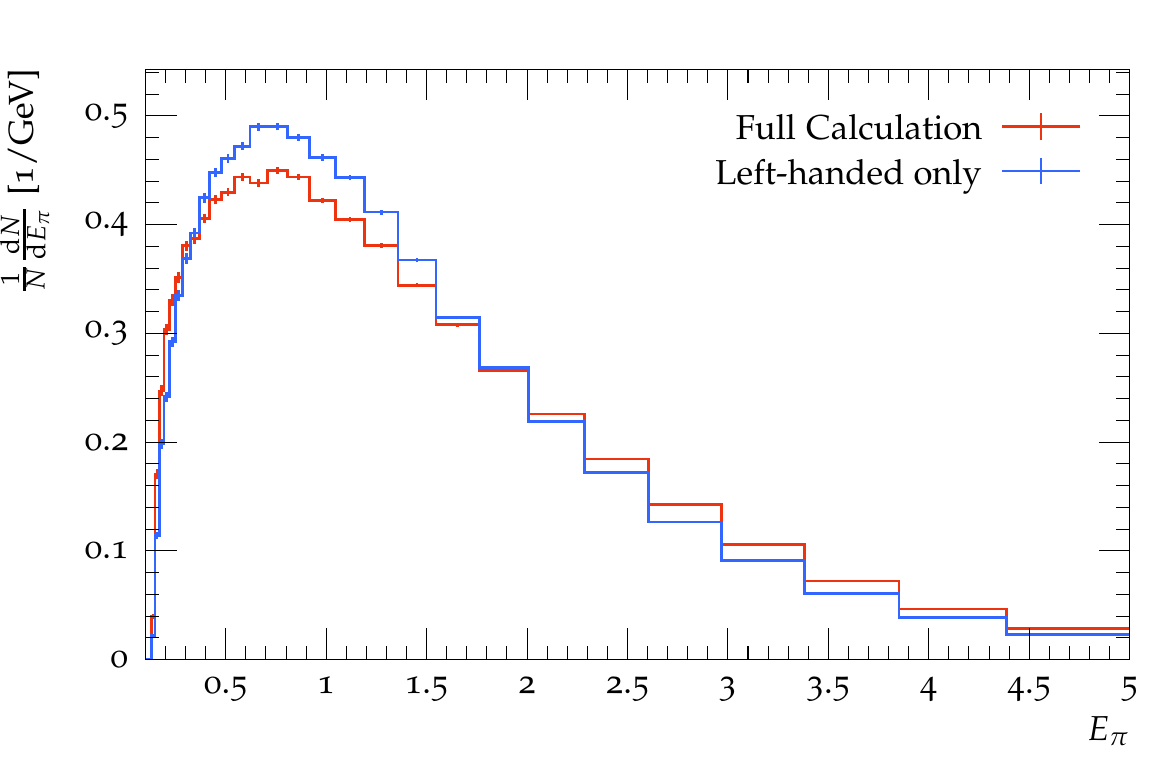} 
    \caption{Energy of the leading pion in $\nu_\tau A \to \tau X$ events, in which all possible decays of the $\tau$
    are included. These results do not include the production of pions from the intranuclear cascade.}
    \label{fig:dune_pedro}
\end{figure}

Finally, we perform the analysis proposed in Ref.~\cite{Machado:2020yxl}. The comparison between the full calculation
and the left-handed polarization assumption is shown in Fig.~\ref{fig:dune_pedro} for the energy of the leading pion.
There is a shift in the energy distribution of the pion when correctly handling the tau polarization, making
the pion slightly harder. The study on the impact of this in the separation from the neutral current background
is left to a future work. Since the final state interactions are turned off in this analysis, the other distributions
given in Ref.~\cite{Machado:2020yxl} would not be accurate. Therefore, they are not included here but will be included
in a detailed study on separating the $\tau$ decays from the background.

\section{Conclusions}

Due to the limited number of identifiable tau neutrino events, the tau neutrino is typically considered the least understood
fundamental particle in the Standard Model. Current and next-generation experiments will collect a large number of tau neutrino events,
opening the door to detailed study of this particle.

One of the most important experiments for studying the tau neutrino will be the DUNE experiment. It will be the only experiment
using accelerator neutrinos for measuring properties of the tau neutrino. At DUNE energies, the quasielastic scattering component
is the dominant contribution. In this energy region, there is an irreducible background from neutral current resonance interactions.
Therefore, it is vital to understand the most optimal way to separate the signal from the background. Traditionally, in neutrino
event generators the outgoing $\tau$ is assumed to be fully left-hand polarized. This assumption is poor for DUNE energies.

In this work, we demonstrate the appropriate way of calculating the polarization of the tau and propagating this information through the full
decay chain within an event generator framework. The simulations were performed with a publicly available version of Achilles
interfaced with Sherpa. For validation, we showed that the distributions for single pion are consistent with Ref.~\cite{Hernandez:2022nmp} for
monochromatic beams. We additionally showed strong shifts in the momentum distributions for the two pion decay channel and found
insignificant shifts (as expected) in the three pion decay channels from the fully left-handed assumption. We also considered the decay in the
leptonic channel, and found a slight shift when correctly handling the polarization.

While the study with monochromatic beams allows for validation of the calculation, all current and future experiments have a
broad spread in the neutrino energies. We therefore investigated the changes in the same distributions integrated over the 
$\tau$-optimized running mode for DUNE. Again we find significant changes from the traditional fully left-handed assumption
in the lepton, single pion, and two pion channel. As expected, there were no significant modifications in the three pion channel.

Finally, while the distributions shown here demonstrate the importance of properly handling the polarization of the tau, they are
not necessarily the optimal variables for separating the tau from the neutral current background. The investigation
of how to optimally separate the charge current tau neutrino interactions from the SM background is left to a
future work.

\section{Acknowledgments}
We thank Joanna Sobczyk and collaborators for insightful discussions. We thank Noemi Rocco, William Jay, and Andr\'e de Gouv\^ea for many useful discussions and for their comments on the manuscript. We thank Pedro Machado for helping with the realistic tau neutrino beams, for many fruitful discussions, and for his comments on the manuscript.
This manuscript has been authored by Fermi Research Alliance, LLC under Contract No. DE-AC02-07CH11359 with the U.S. Department of Energy, Office of Science, Office of High Energy Physics.

\appendix

\bibliography{biblio}
\end{document}